\documentclass[11 pt, letterpaper]{article}
\usepackage{graphicx}
\usepackage{epstopdf, epsfig}

\usepackage{amsmath, amssymb, latexsym, marvosym}
\usepackage{natbib}  
\usepackage{color}  

\usepackage[letterpaper, margin=0.75in]{geometry}

\newcommand{\ie}{\textit{i.e.} }
\newcommand{\be}{\begin{equation}}
\newcommand{\ee}{\end{equation}}
\newcommand{\ba}{\begin{align}}
\newcommand{\ea}{\end{align}}
  
\newcommand{\bs}{\boldsymbol}
   
   \newcommand{\x}{\textbf{x}}

\renewcommand{\u}{\textbf{u}}
\renewcommand{\v}{\textbf{v}}
\newcommand{\bc}{\bar{c}}
\newcommand{\bu}{\bar{u}}
\newcommand{\bv}{\bar{v}}
\newcommand{\bw}{\bar{w}}
\newcommand{\bxi}{\bs\xi}
\newcommand{\btxi}{\bs{\tilde\xi}}
\newcommand{\bbxi}{\bs{\bar\xi}}
\newcommand{\bteta}{\bs{\tilde\eta}}
\newcommand{\bbeta}{\bs{\bar\eta}}
\newcommand{\ub}{\bs{\bar{\textbf{u}}}}
\newcommand{\vb}{\bs{\bar{\textbf{v}}}}

\newcommand{\vt}{\bs{\tilde{\textbf{v}}}}

\newcommand{\p}{\partial}
\newcommand{\pt}{\partial t}

\newcommand{\TAU}{\bs{\tau}}

\renewcommand{\ss}{\scriptstyle}
\renewcommand{\O}{{\cal O}}

\renewcommand{\[}{\left[}
\renewcommand{\]}{\right]}
\renewcommand{\(}{\left(}
\renewcommand{\)}{\right)}
\newcommand{\half}{\tfrac{1}{2}}

\newcommand{\ra}{\rightarrow}
\newcommand{\alfa}{\alpha}
\newcommand{\eps}{\epsilon}

\newcommand{\T}{\intercal} % transpose symbol

\newcommand{\mediumint}{\text{\scriptsize $\int$}}
\newcommand{\mediumiint}{\text{\scriptsize $\iint$}}
\newcommand{\mediumiiint}{\text{\scriptsize $\iiint$}}

\def\Xint#1{\mathchoice
{\XXint\displaystyle\textstyle{#1}}%
{\XXint\textstyle\scriptstyle{#1}}%
{\XXint\scriptstyle\scriptscriptstyle{#1}}%
{\XXint\scriptscriptstyle\scriptscriptstyle{#1}}%
\!\int}
\def\XXint#1#2#3{{\setbox0=\hbox{$#1{#2#3}{\int}$ }
\vcenter{\hbox{$#2#3$ }}\kern-.6\wd0}}

\def\dashint{\Xint-}
\def\XXinttext#1#2#3{{\setbox0=\hbox{$#1{#2#3}{\int}$ }
\vcenter{\hbox{$#2#3$ }}\kern-.85\wd0}}

\begin{document}

\renewcommand{\baselinestretch}{1.1}

\begin{center}
{\LARGE\bf
Turbulence Modeling via the  Fractional Laplacian \\[1ex]
} 

{\Large
Brenden P. Epps and Benoit Cushman-Roisin \\[1ex]
}

Thayer School of Engineering, Dartmouth College, Hanover, NH 03755, U.S.A.\\[1 em]

{\color{red}
This draft was submitted to the Journal of Fluid Mechanics on 28-November 2017 for peer review.\\[1 em] }

\end{center}

\begin{abstract}
Herein, we provide the first ever derivation of the {\it fractional Laplacian} operator as a means to represent the mean friction force arising in a turbulent flow: 
\begin{equation}
\rho \frac{D\ub}{Dt}
= -\nabla p + \mu_\alfa \nabla^2\ub + \rho C_\alfa \dashint\!\!\!\dashint\!\!\!\dashint_{\!-\infty}^\infty
    \frac{  \ub{\ss(t,\x')} - \ub{\ss(t,\x)}  }{|\x'-\x|^{\alfa+3}} \,d\x' 
\label{eq_momentum_0}
\end{equation}
where $\ub{\ss(t,\x)}$ is the ensemble-averaged velocity field, 
$\mu_\alfa$ is an enhanced molecular viscosity, and $C_\alfa$ is a turbulent mixing coefficient (with units (length)$^\alfa$/(time)). 
The derivation is grounded in Boltzmann kinetic theory, which presumes an equilibrium probability distribution $f_\alpha^{eq}(t,{\bf x},{\bf u})$ of particle speeds.  While historically $f_\alpha^{eq}$ has been assumed to be the {\it Maxwell-Boltzmann (normal) distribution}, we show that any member of the family of {\it L\'evy $\alpha$-stable distributions} is a suitable alternative, with parameter $\alfa$ selecting the distribution.
  If $\alfa=2$, then $f^{eq}_\alfa$ is the {\it Maxwell-Boltzmann distribution}, with large particle speeds very unlikely, and \eqref{eq_momentum_0} reverts to the Navier-Stokes equation (with $\mu_\alfa = \mu$ and $C_\alfa = 0$).  If $0 < \alfa < 2$, then $f^{eq}_\alfa$ is a {\it L\'evy $\alpha$-stable distribution}, with ``heavy tails'' that permit large velocity fluctuations, as in turbulence. 
For shear turbulent flows, the choice of $\alpha = 1$ ({\it Cauchy distribution} for $f_\alpha^{eq}$) leads to the logarithmic velocity profile known as the {\it Law of the Wall}.  The only restrictions in the present derivation are assumptions of a low Knudsen number,  incompressible flow, and isothermal flow.  With these assumptions, we show how 
the Boltzmann kinetic equation and {\it L\'evy $\alpha$-stable distributions} lead  
to Equation \eqref{eq_momentum_0}.  We also present examples of 1D Couette flow and 2D boundary layer flow, and we discuss turbulent transport within this kinetic theory framework.  This work lays out a new framework for turbulence modeling that may lead to new fundamental understanding of turbulent flows.
\end{abstract}

\section{Introduction}

\subsection{Literature Review}

While it is widely accepted that the Navier-Stokes equations describe turbulent fluid flows, direct numerical simulation of these equations for high-Reynolds-number engineering flows ($Re \gtrsim \O(10^6))$ would take lifetimes to compute.  Thus, a wide array of turbulence models have been proposed  \citep{Tennekes1972,Pope2000,Wilcox2006,McDonough2007}.  However, any path that follows  Osborne Reynolds' (1895) \nocite{Reynolds1895} method of averaging the governing equations, and then solving, leads to the {\it closure problem}, with more unknowns than equations.  
Our approach is opposite of Reynolds'; starting with the Boltzmann equation from kinetic theory, we first solve and then ensemble average.  The result of our analysis is Equation \eqref{eq_momentum_0}, which governs the ensemble-averaged flowfield and uses the fractional Laplacian to represent turbulent friction.  It should be emphasized that 
this operator is not chosen {\it ad hoc} but rather is derived from the first principles of Boltzmann kinetic theory and the statistics of turbulent transport.  
Our premise is that turbulence is a {\it non-local} phenomenon, statistically bringing together distant fluid particles and causing exchanges of mass and momentum from distant parts of the flowfield. Thus, this literature review focuses on three {key ideas}: 
(i) the description of {\it anomalous diffusion} using {\it L\'evy $\alpha$-stable distributions};
(ii) {\it non-local} turbulence closure models, including {\it fractional derivatives};
 and
(iii) the {\it kinetic theory} framework for turbulence modeling.

\begin{figure}
\begin{center}
\includegraphics[width=0.9\textwidth]{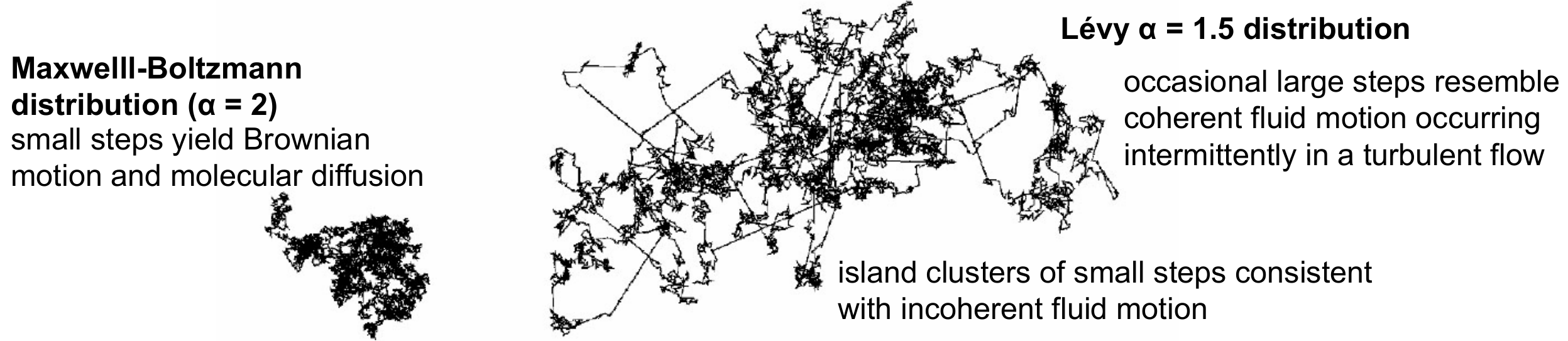}
\caption{Fluid particle trajectories with steps drawn from a Maxwell-Boltzmann distribution (left) and L\'evy $\alfa=\frac{3}{2}$ distribution (right).  Both trajectories have 7000 steps. 
  Figure from [Chechkin {\it et al.}, 2008], with annotations.}
\label{fig_trajectories}
\end{center}
\end{figure}

Turbulence exhibits {\it anomalous diffusion}, which means that diffusion occurs over distances $\xi$ that scale by time to a power other than one half, $\xi \sim \O( t^{1/2})$, which is the scaling for molecular diffusion \citep{Zaslavsky2002}.   
For example, air pollution dispersed from a point source will disperse in a patch that grows linearly with time - {\it i.e.\  super-diffusion} with length scale $\xi \sim \O(t)$ \citep{Kampf2016}. 
 This super-diffusion can be accurately modeled using a fractional Laplacian for the diffusion operator (in lieu of the regular Laplacian as in the usual $\O(t^{1/2})$ Fickian diffusion) \citep{CushmanRoisin2008,CushmanRoisin2013}.
In another famous example, \citet{Richardson1926} observed that the root-mean-square distance $\langle\xi\rangle$ separating fluid particles initially near one another in a turbulent atmosphere scales as $\langle\xi\rangle \sim t^{3/2}$.  \cite{Shlesinger1987} formally showed that Richardson's law could be derived assuming the turbulence followed the $\alfa = \frac{2}{3}$ L\'evy distribution.  In their derivation, they assumed that the trajectory of a fluid particle may follow a {\it L\'evy walk}, characterized by occasional large steps due to coherent fluid motion (see {\bf Figure \ref{fig_trajectories}}).  L\'evy walks are random walks where the particle moves with constant velocity for random periods of time, instantly choosing another random velocity at each turning point (collision with another particle within the context of Boltzmann kinetics).

{\it L\'evy $\alpha$-stable distributions} (for $0 < \alfa < 2$) 
are probability distributions that describe fluctuating processes characterized by large bursts or outliers, such as turbulence \citep{Levy1937, Chechkin2008, Nolan2017, Shintani2017}.   
For example,  wind speeds have been shown to be L\'evy-distributed with $1.5 \lesssim \alfa \lesssim 1.72$ \citep{Boettcher2003,Metzler2009,Blackledge2011}.  L\'evy-distributed velocity fluctuations with $\alfa \approx 1$ have been observed in studies such as:
turbulent pipe flow behind a grid \citep{Tong1988, Onuki1988};
flowfield induced by a large number of point vortices \citep{Min1996}; and
Couette flow between parallel rotating disks \citep{Mordant2001}.

A number of {\it non-local turbulence models} have been explored in the literature, including the fractional Laplacian.   
Without derivation, W. Chen (2006) \nocite{WChen2006} speculated that {\it inertial-range} turbulence could be modeled by the {fractional Laplacian} with $\alfa = 2/3$. 
 Recently, \citet{Churbanov2016} used the fractional Laplacian to model turbulent flow in a rectangular duct, with $\alfa=1/4$ (chosen {\it ad hoc}) showing reasonable agreement with experiments.
Similarly, \citet{Xu2017} analyzed plane Poiseuille flow with encouraging results.
Modification of Navier-Stokes with a fractional time derivative has also been proposed without derivation \citep{ElShahed2004, Kumar2015}.
However it is important to note that the literature neither offers a rigorous derivation of the fractional Laplacian as a turbulence model, nor offers a theory as to how to choose the fractional order $\alfa$.

A number of  non-local diffusion operators have also been proposed.
Based on phenomenological arguments, \cite{Schumer2003,Schumer2009} model advection-dispersion using the fractional Laplacian. 
  The forms of other diffusion operators stem from the fact that the Fourier transform of  the regular Laplacian  $\mu \nabla^2 u$
 is $- \mu |k|^2 \hat{u}$, where $k$ is the wavenumber.\footnote{The Fourier transform of the {\it fractional Laplacian} is $-|k|^\alfa \hat{u}$.}
Thus, \citet{Berkowicz1980} proposed $- \mu{\ss(k)} |k|^2 \hat{u}$ , where $\mu{\ss(k)}$ is a an {\it ad hoc} function of wavenumber $k$.    
Following suit, several workers proposed various forms for $\mu{\ss(k)}$, differing essentially in the manner by which they construct $\mu{\ss(k)}$ and the number of tunable parameters that it contains \citep{Fiedler1984, Stull1984,Stull1993,Nakayama1995}. 
One model of note is that of \citet{CushmanRoisin2006}:
$
F{\ss(z)} = \frac{\rho A}{\pi} \int_{0}^\infty \frac{|u{\ss(z')}-u{\ss(z)}| (u{\ss(z')}-u{\ss(z)})}{(z'-z)^2} \, dz' .
$  
For wall-bounded shear flows, the solution of $F{\ss(z)} = 0$ is the logarithmic profile $u{\ss(z)} = \ln{\ss(z)}$, in agreement with the {\it Law of the Wall}.

Efforts have been made to model turbulence using {\it Boltzmann kinetic theory}.  
Several authors have pursued numerical solution of the {\it Boltzmann equation}  with different 
{\it collision operators} (right hand sides), including:
non-local collision operator \citep{Hayot1996};
fractional time derivative \citep{Baule2006};
orthogonal projector \citep{Degond2002}; or 
an integral forcing term \citep{Tsuge1976, Srinivasan1966}.
In order to describe Reynolds stresses using kinetic theory, \citet{Girimaji2007}  decomposes the Boltzmann equation into filtered and unresolved parts, and he shows direct correspondence between the resulting kinetic model and the Reynolds stresses. 

A noteworthy series of papers in this journal is that of H.\ Chen and colleagues, who explored the Boltzmann equation with Maxwell-Boltzmann $f^{eq}$ but large collision time $\tau$.  
\citet{Chen2004}  extended the {\it Chapman-Enskog expansion} $f = f^{eq} + \eps f^{(1)} + \dots$ to include the second order term $\eps^2 f^{(2)}$ and found that using a finite collision time $\tau$ leads to memory effects and nonlinear constitutive relations consistent with turbulence.  
\citet{Chen2007,Chen2010}  provided the exact solution of the Boltzmann equation \eqref{eq_f_general_soln} and used it to evaluate the stress tensor \eqref{eq_sigma_def}.  Considering Maxwell-Boltzmann $f^{eq}_\alfa$ and finite $\tau$, they determined an integro-differential equation that reduces to the Navier-Stokes equations in the limit $\tau\ra 0$.  
\citet{Chen2013} applied their model (finite $\tau$, Gaussian $f^{eq}_\alfa$) to Couette flow  and found that (i) for small $\tau$, their theory replicated a linear velocity profile consistent with Navier-Stokes, as expected; but (ii) for large $\tau$, the velocity profile was still mostly linear but exhibited slip along the walls.  The latter is not in agreement with data   
\citep{Robertson1970}, which reveal an S-shape velocity profile without slip along the walls.

%----------------------------------------------------------------------------------------------------
%----------------------------------------------------------------------------------------------------
%----------------------------------------------------------------------------------------------------
%----------------------------------------------------------------------------------------------------
\subsection{Key Ideas and Assumptions}
Several elements distinguish this work from the literature:
 While the fractional Laplacian has been used previously, we provide the first rigorous derivation of the fractional Laplacian for the representation of turbulence. 
 Moreover, we provide the first theory rigorously justifying the choice of parameter $\alfa$: The appropriate $\alfa$ is the one that corresponds to the scaling $\delta{\ss(t)} \sim t^\frac{1}{\alfa}$ of the observed macro-scale transport.
  Our derivation is rooted in the Boltzmann equation and L\'evy statistics, two ideas that have been used to describe turbulence for some time but never combined as they are herein.  
Finally, while H.\ Chen and colleagues have considered a large $\tau$ and {\it Maxwell-Boltzmann} $f^{eq}_\alfa$, this paper considers small $\tau$ and other L\'evy  $\alfa$-stable $f^{eq}_\alfa$.

The key ideas that lead to Equation \eqref{eq_momentum_0} are as follows:
\begin{itemize}
\item  Our foundation is {\it Boltzmann kinetic theory}, which has been shown to lead to the Navier-Stokes equations if the distribution of molecular speeds $f^{eq}_\alfa$ is assumed to be a {\it Maxwell-Boltzmann} ({\it normal}) distribution.

\item For reasons discussed in \S\ref{sec_transport} and \S\ref{sec_scales}, we consider the broader family of {\it L\'evy $\alfa$-stable distributions}, of which the Maxwell-Boltzmann distribution is but one member.  
 
\item The key algebra ``trick'' is to separately consider ``small'' and ``large'' particle displacements while evaluating the stress field.

\end{itemize}

The only restrictions in the present derivation are assumptions of low Knudsen number, incompressible flow, and isothermal flow.  With these assumptions, we show how the Boltzmann equation and L\'evy $\alfa$-stable distributions lead directly to Equation \eqref{eq_momentum_0}.

%----------------------------------------------------------------------------------------------------
%----------------------------------------------------------------------------------------------------
%----------------------------------------------------------------------------------------------------
%----------------------------------------------------------------------------------------------------
\subsection{Organization of this Article}
In \S\ref{sec_transport} we develop a number of key ideas needed for \S\ref{sec_derivation} by considering turbulent transport of passive scalars.
In \S\ref{sec_derivation}, we derive the fractional Laplacian as a model for the ensemble-averaged friction force arising in a turbulent flow.   
Section \S\ref{sec_examples} provides properties of the fractional Laplacian operator, as well as examples of 1D Couette flow and 2D boundary layer flow.  Finally, conclusions are offered in \S\ref{sec_conclusions}, and appendices provide supplemental information.

%\clearpage
%----------------------------------------------------------------------------------------------------
%----------------------------------------------------------------------------------------------------
%----------------------------------------------------------------------------------------------------
%----------------------------------------------------------------------------------------------------
%----------------------------------------------------------------------------------------------------
%----------------------------------------------------------------------------------------------------
%----------------------------------------------------------------------------------------------------
%----------------------------------------------------------------------------------------------------%----------------------------------------------------------------------------------------------------
%----------------------------------------------------------------------------------------------------
%----------------------------------------------------------------------------------------------------
%----------------------------------------------------------------------------------------------------
\section{Turbulent Transport}
\label{sec_transport}

In this section, we consider the transport of passive scalars in a turbulent flow.  Through this warm-up problem, we  introduce several ideas, including 
the probability and phase-space fundamentals needed for the more complex analysis of momentum transport in \S\ref{sec_derivation}.  We derive the requirement that the probability distribution describing particle velocities are {\it stable} and then show that this stability constraint is satisfied by the {\it L\'evy $\alfa$-stable distributions}.  Further, we develop the hypothesis that these probability distributions are self-similar and that a similarity variable can be defined that links the macro-scale transport with the micro-scale molecular motions. Finally, we conclude that turbulent diffusion can be described by a fractional Laplacian operator, paving the way for the momentum analysis in \S\ref{sec_derivation}.

Our premise is to consider transport of a conserved quantity $c{\ss(t,\x;\u)}$ being advected by a random flow field, $\u$.  
Let $c{\ss(t,\x;\u)}$ be a passive scalar quantity, such as the concentration of some species (of particles with velocity $\u$).  
The 3D transport equation is
\begin{equation}
     \frac{\p c}{\pt} +  \u \cdot \nabla c = 0
\label{eq_3D_governing_eqn}
\end{equation}
with zero right hand side to indicate no changes to $c$ due to sources/sinks. 
Random velocity $\u = [u,v,w]$ is a 3D random variable governed by probability density function $p_\u{\ss(\u)}$, which denotes the joint pdf of $p_\u{\ss(u,v,w)}$.  At this point, we require two conditions on $p_\u{\ss(\u)}$:
\begin{itemize}
\item {\it Normalization:}
\be
\mediumiiint_{\!\!\!-\infty}^\infty \, p_\u{\ss(\u)} d\u = 1  ~.
\label{eq_pu_normalization}
\ee

\item {\it Isotropy:}  That is, $p_\u(\u)$ is a function of the magnitude of the velocity deviation $|\u-\ub|$, where $\ub{\ss(t,\x)}$ is the mean flow speed. 

\end{itemize}

The observable average, $\bc$, then is 
\be
\bc{\ss(t,\x)} = \mediumiiint_{\!\!\!-\infty}^{\infty} ~ c{\ss(t,\x; \u)} p_\u{\ss(\u)} \, d\u  ~.
\label{eq_c_avg}
\ee

We would like to determine the ensemble-averaged behavior of \eqref{eq_3D_governing_eqn} using the probability/phase-space framework.  Our general strategy is opposite of Osborne \citet{Reynolds1895}: While he averaged the equations and then solved, we solve equation \eqref{eq_3D_governing_eqn} and then ensemble average.   For any single realization of random velocity $\u$ over time interval $\delta t$, the solution of \eqref{eq_3D_governing_eqn} is 
\begin{equation}
c(t + \delta t, \x;\u) = c(t,\x - \u \delta t; \u) ~,
\label{eq_1D_solution}
\end{equation}
which states that $c$ is merely advected by the flow. 

The ensemble-averaged behavior of \eqref{eq_3D_governing_eqn} can be deduced as follows.   First, construct the ensemble-averaged time derivative consistent with the rules of calculus:
\begin{equation}
\frac{\p \bar{c}}{\pt} \equiv \lim\limits_{\delta t \ra 0} \frac{ \bar{c}(t+\delta t,\x) - \bar{c}(t,\x) }{\delta t}  ~.
\label{eq_1D_dcdt}
\end{equation}
Using \eqref{eq_c_avg} and \eqref{eq_1D_solution} to evaluate  \eqref{eq_1D_dcdt}, we can write 
\begin{equation}
\frac{\p \bar{c}}{\pt} 
= \lim\limits_{\delta t \ra 0} \iiint\limits_{\!\!\!-\infty}^{~~~\infty} ~  \frac{ c(t,\x-\u\,\delta t;\u) - c(t,\x;\u) }{\delta t}  p_\u{\ss(\u)} \, d\u  ~.
\label{eq_1D_dcdt_0}
\end{equation}
To proceed, we need to constrain the probability distribution $p_\u{\ss(\u)}$.

%------------------------------------------------------------------
\subsection{Stability Constraint}

Note that as $\delta t \ra 0$ in \eqref{eq_1D_dcdt_0}, the probability $p_\u{\ss(\u)}$ must remain non-trivial.  This requires that $p_\u{\ss(\u)}$ be a {\it stable distribution}.  
A probability distribution is {\it stable} if the shape of the distribution is preserved under addition.  That is, if  $X_1$ and $X_2$ are two independent random variables drawn from a stable distribution, then their sum $X_1 + X_2$ also follows that distribution (up to scale and shift) \citep{Nolan2017}.

In order to better expose this requirement, 
switch from velocity to the corresponding displacement: 
\be
\bxi = \u \, \delta t ~.
\label{eq_xi_to_udt}
\ee
These displacements are governed by a pdf such that 
$
p_{\bxi}{\ss(\bxi ; \delta t)} \, d\bxi = p_\u{\ss(\u)} d\u
$ 
where $d\bxi = d\xi_1 \, d\xi_2 \, d\xi_3 = (u \, \delta t)(v \, \delta t)(w \, \delta t) = (\delta t)^3 d\u$.  Thus,
\be
p_{\bxi}{\ss(\bxi ; \delta t)}   = p_\u{\ss(\u \,=\, \bxi/\delta t)} / (\delta t)^3 ~.
\label{eq_pxi2pu}
\ee
We simultaneously require that as $\delta t \ra 0$, $p_{\bxi}{\ss(\bxi ; \delta t)}$ becomes a Dirac function $\delta(\bxi - \bbxi)$, while  $p_{\bxi}{\ss(\bxi ; \delta t)} / \delta t$ remains finite.
Together, these requirements produce streaming to the mean displacement $\bbxi$ and finite diffusion about that location.  
In order for these constraints to be simultaneously met,  $p_{\bxi}{\ss(\bxi ; \delta t)}$ must be a {\it stable distribution}.  

Realizing that the shift $\bbxi = \ub \, \delta t$ represents the mean motion, we simplify the exposition by introducing the displacement deviation $\btxi = \bxi - \bs{\bar{\xi}}$ and corresponding probability distribution, $p_{\btxi}{\ss(\btxi ; \delta t)} = p_{\bs\xi}{\ss(\bxi \,=\, \bbxi + \btxi ; \delta t)}$.  Then, distribution $p_{\btxi}{\ss(\btxi ; \delta t)}$ must be a stable distribution with zero shift.

The simplest way of ensuring that $p_{\btxi}{\ss(\btxi ; \delta t)}$ be a stable distribution is to require that the displacement deviation $\btxi$ in time interval $\delta t$ follows the same pdf as the displacement in half of that time interval.  In other words, the probability of jump $\btxi$ in time $\delta t$ is given by the convolution of the probabilities of all possible intermediate jumps  $\btxi'$ and then complementary jumps $\btxi-\btxi'$ made in successive half intervals $\delta t / 2$.\footnotemark
\begin{equation}
p_{\btxi}(\btxi;\delta t) = \iiint\limits_{\!\!\!-\infty}^{~~~\infty} p_{\btxi}(\btxi';\delta t/2) p_{\btxi}(\btxi-\btxi';\delta t/2) \, d\btxi'  ~.
\label{eq_divisibility}
\end{equation}
If the time interval can be halved, it can be halved once more, and so forth {\it ad infinitum} to reach the limit $\delta t \ra 0$.

\footnotetext{
Properly stated, the stability constraint \eqref{eq_divisibility} should hold for any initial jump displacement $\btxi'$ in time lapse $\delta t'$, followed by complementary $\btxi-\btxi'$ in $\delta t - \delta t'$.  It is straightforward to show that solutions to \eqref{eq_divisibility} also satisfy this more general requirement.
}

For applicability across systems of different sizes and flow fields, there must exist a general formulation in which the structure of the pdf remains the same irrespective of the spatial size and strength of the flowfield.  This necessitates the existence of a dimensionless variable with a canonical pdf.  This in turn necessitates the reliance on a dimensional quantity that can be used for scaling.  Table \ref{tab_similarity} lists three such possible scalings.
In general, we invoke the existence of a dimensional quantity, $q$, with dimensions (length)$^\alfa$ / (time), from which we can form the scaled displacement (similarity variable) $\bs{\tilde\eta}$ from the dimensional displacement $\btxi$:
\be
\bs{\tilde\eta} \equiv \btxi / (q \,\delta t)^{1/\alfa} ~.
\label{eq_eta_def}
\ee

We assert that \eqref{eq_eta_def} holds in general, with parameter $\alfa$ describing the temporal scaling of the turbulent transport. 
This hypothesis is well justified by experimental evidence for $\alfa = 2$, 1, and 2/3.    
For laminar diffusion $\alfa = 2$, momentum transport scales as $\xi \sim (\nu \, \delta t)^\frac{1}{2}$.  For wall bounded turbulent flows $\alfa = 1$,  turbulent transport scales as $\xi \sim u_* \, \delta t$, where $u_* \equiv \sqrt{\tau_w/\rho}$ is the {\it friction velocity}.\footnote{The relevant physical constant is the wall stress, $\tau_w$, which is repackaged as a friction velocity, $u_* \equiv \sqrt{\tau_w/\rho}$ in order to have the correct units: (length)$^1$/(time).}  Richardson's (1926) observations showed turbulent dispersion scaling as $\xi \sim t^{3/2}$, corresponding to $\alfa = 2/3$; from the Kolmogorov energy cascade of inertial turbulence, the pertinent dimensional quantity is the energy dissipation rate $\epsilon$, giving $\xi \sim (\epsilon^{1/3} \delta t)^{3/2}$ consistent with Richardson's observations.

\begin{table}
\begin{center}
\caption{Physical parameters and similarity variables associated with molecular diffusion, shear turbulence, and inertial range  turbulence.}\vspace{1 ex} 
\label{tab_similarity}
\begin{tabular}{lcccc}
{\it physical parameter} 			& {\it ~~units~~} 				& {\it ~similarity variable~}					&  {\it exponent}	& {\it distribution} \\[1 ex]
kinematic viscosity, $\nu$			& $L^2/T$ 	& $\bs\eta = \bxi/(\nu \,\delta t)^\frac{1}{2}$ 		& $\alfa = 2$ & Maxwell-Boltzmann \\[1 ex]
friction velocity, $u_*$					& $L/T$ 				& $\bs\eta = \bxi/(u_* \,\delta t)$ 			& $\alfa = 1$ 	& Cauchy \\[1 ex]
energy dissipation rate, $\epsilon^\frac{1}{3}$& $L^\frac{2}{3}/T$ 	& $\bs\eta = \bxi/(\epsilon^\frac{1}{3} \,\delta t)^\frac{3}{2}$ & $\alfa = \tfrac{2}{3}$ & other L\'evy \\[1 ex]
general, $q$  & $L^\alfa/T$ 	& $\bs\eta = \bxi/(q \,\delta t)^{1/\alfa}$ & $\alfa$ & other L\'evy 
\end{tabular}
\end{center}
\end{table}%

In order to solve Equation \eqref{eq_divisibility}, we must recast it in terms of a universal probability density function $p_{\bteta}{\ss(\bteta)}$ that governs the scaled displacement $\bteta$.
So we define $p_{\bs\eta}{\ss(\bs\eta)}$ such that 
$
p_{\bs\eta}{\ss(\bs\eta)} \, d\bs\eta = p_{\bxi}{\ss(\bxi; \delta t)} \, d\bxi
$
with
$d\bs\eta = d\eta_1 \, d\eta_2 \, d\eta_3 = d\xi_1 \, d\xi_2 \, d\xi_3 / (q \, \delta t)^{3/\alfa}  = d\bxi / (q \, \delta t)^{3/\alfa}$. Thus 
$d\bxi = (q \, \delta t)^{3/\alfa} d\bs\eta $, and 
\begin{align}
 p_{\bxi}{\ss(\bxi ; \delta t)} &= p_{\bs\eta}{\ss(\bs\eta \,=\, \bxi / (q \,\delta t)^{1/\alfa} )} /  (q \, \delta t)^{3/\alfa} ~. \label{eq_G2g}  
\end{align}
To convert $p_{\btxi}{\ss(\btxi' ; \delta t/2)}$, we need to determine the scaled parameter that corresponds to a jump of 
$\btxi'$ in half time interval $\delta t/2$; this value is 
$\btxi' / (q \, \delta t/2)^{1/\alfa} 
= 2^{1/\alfa} \btxi' / (q \, \delta t)^{1/\alfa}
= 2^{1/\alfa} \bteta'$.
Thus, $p_{\btxi}{\ss(\btxi' ; \delta t/2)} 
= p_{\bteta}{\ss( 2^{1/\alfa}  \bteta' )} /  (q \, \delta t/2)^{3/\alfa}$.    
Using these conversions, Equation \eqref{eq_divisibility} can be written as
\be
\frac{p_{\bteta}{\ss(\bteta)} }{ (q \, \delta t)^{3/\alfa} }   
 = \iiint\limits_{\!\!\!-\infty}^{~~~\infty} 
 \frac{p_{\bteta}{\ss(2^{1/\alfa} \bteta')} }{ (q \, \delta t/2)^{3/\alfa} }  
 \frac{p_{\bteta}{\ss(2^{1/\alfa} (\bteta -\bteta'))} }{ (q \, \delta t/2)^{3/\alfa} } \, d\bteta'  (q \, \delta t)^{3/\alfa}  ~,
\label{eq_divisibility_G}
\ee
which simplifies to 
\be
p_{\bteta}{\ss(\bteta)}
 = \iiint\limits_{\!\!\!-\infty}^{~~~\infty} 
 2^{6/\alfa} \, p_{\bteta}{\ss(2^{1/\alfa} \bteta')}  \, p_{\bteta}{\ss(2^{1/\alfa} (\bteta -\bteta'))}
\, d\bteta'   ~.
\label{eq_divisibility_G2}
\ee
We can put \eqref{eq_divisibility_G2} into a more friendly form by setting $\bs{\check\eta} = 2^{1/\alfa} \bteta$ such that $d\bs{\check\eta}' = 2^{3/\alfa} d\bteta'$.  Upon doing so and immediately replacing the checks with tildes, equation  \eqref{eq_divisibility_G2}  
simplifies to the {\it stability constraint equation}:
\be
2^{-3/\alfa} p_{\bteta}{\ss(2^{-1/\alfa} \bteta)}
 = \iiint\limits_{\!\!\!-\infty}^{~~~\infty} 
\, p_{\bteta}{\ss(\bteta')}  \, p_{\bteta}{\ss(\bteta -\bteta')}
\,  d\bteta'   ~.
\label{eq_divisibility_soln2}
\ee

Equation \eqref{eq_divisibility_soln2} provides a constraint on which probability distributions are admissible.  Solutions to \eqref{eq_divisibility_soln2} may be used to model turbulent flows.

\subsection{L\'evy $\alfa$-stable Distributions}
In this subsection, we show that the solution to the stability constraint equation \eqref{eq_divisibility_soln2} is the {\it L\'evy $\alfa$-stable distribution}.  The derivation is straightforward, taking the Fourier transform of \eqref{eq_divisibility_soln2}, solving in Fourier space, and inverse Fourier transforming the solution.

Herein, we adopt the unitary form of the Fourier transform (in $3$-dimensional space):
\begin{align}
\text{transform:}    \quad       \quad
{\cal F}\{ p_{\bteta}{\ss(\bteta)} \}{\ss({\bf k})} \equiv
\widehat{p_{\bteta}}{\ss({\bf k})} &\equiv \frac{1}{(2\pi)^{3/2}} \int_{\mathbb{R}^3} \, p_{\bteta}{\ss(\bteta)} e^{-i{\bf k}\cdot\bteta} \, d\bteta ~; \\
\text{inverse transform:}  \quad  
{\cal F}^{-1}\{ \widehat{p_{\bteta}}{\ss({\bf k})}  \}{\ss(\bteta)} \equiv 
p_{\bteta}{\ss(\bteta)} &\equiv \frac{1}{(2\pi)^{3/2}} \int_{\mathbb{R}^3} \, \widehat{p_{\bteta}}{\ss({\bf k})}  e^{ i{\bf k}\cdot\bteta} \, d{\bf k} ~.
\end{align}

In order to evaluate the Fourier transform of \eqref{eq_divisibility_soln2}, we will use the scale property:
${\cal F}\{ p_{\bteta}{\ss(a \bteta)} \}{\ss({\bf k})} = \frac{1}{|a|^3} \widehat{p_{\bteta}}{\ss({\bf k}/{a})}$
and the {\it convolution theorem}:
${\cal F}\{ (f \otimes g){\ss(\bteta)} \}{\ss({\bf k}) }= (2\pi)^\frac{3}{2} \hat{f}{\ss({\bf k})}\hat{g}{\ss({\bf k})}$. 
Thus, the Fourier transform of \eqref{eq_divisibility_soln2} is
\begin{equation}
\widehat{p_{\bteta}}{\ss(2^{1/\alfa} {\bf k})} = (2\pi)^\frac{3}{2} \( \widehat{p_{\bteta}}{\ss({\bf k})} \)^2 ~,
\label{eq_divisibility_soln3}
\end{equation}
of which the solution is
\begin{equation}
\widehat{p_{\bteta}}{\ss({\bf k})} =  \frac{1}{(2\pi)^{3/2}} \, e^{-|\gamma {\bf k}|^\alfa}  ~,
\label{eq_divisibility_soln4}
\end{equation}
where $\gamma$ is a free parameter.
The inverse Fourier transform of \eqref{eq_divisibility_soln4} is
\begin{equation}
 p_{\bteta}{\ss(\bteta)} = 
 \frac{1}{(2\pi)^3} \int_{\mathbb{R}^3} \,  e^{-|\gamma {\bf k}|^\alfa} e^{ i{\bf k}\cdot\bteta} \, d{\bf k}  
  ~,
\label{eq_divisibility_soln_5}
\end{equation}
which is the {\it multivariate L\'evy $\alfa$-stable distribution} with scale $\gamma$ but no skew or shift. Equation \eqref{eq_divisibility_soln_5} can alternatively be written as
\begin{equation}
 p_{\bteta}{\ss(\bteta)} 
 =
  \frac{1}{\gamma^3} \frac{1}{(2\pi)^3} \int_{\mathbb{R}^3} \,  e^{-|{\bf k}|^\alfa} e^{ i{\bf k}\cdot\bteta/\gamma} \, d{\bf k}
  ~.
\label{eq_divisibility_soln_5b}
\end{equation}

%------------------------------------------------
\subsection{Ensemble-Averaged Transport Behavior}

Returning to the problem of turbulent transport, the ensemble-averaged transport, equation \eqref{eq_1D_dcdt_0}, can be put into similarity form upon substitution $p_\u{\ss(\u)} \, d\u = 
p_{\bs\eta}{\ss(\bs\eta)} \, d\bs\eta$ and $\u \,\delta t = \bxi = (q \, \delta t)^{1/\alfa} \bs\eta$:
\begin{equation}
\frac{\p \bar{c}}{\pt} 
= \lim\limits_{\delta t \ra 0} \iiint\limits_{\!\!\!-\infty}^{~~~\infty} ~  \frac{ c(t,\x- (q \, \delta t)^{1/\alfa} \bs\eta; \bs\eta) - c(t,\x;\bs\eta) }{\delta t}  \, p_{\bs\eta}{\ss(\bs\eta)} \, d\bs\eta  ~.
\label{eq_1D_dcdt_1}
\end{equation}
Upon specification of $q$, $\alfa$, and $p_{\bs\eta}{\ss(\bs\eta)} = p_{\bteta}{\ss(\bteta \,=\, \bs\eta - \bbeta)}$, we can evaluate \eqref{eq_1D_dcdt_1}.

%------------------------------------------------
\subsubsection{Ensemble-averaged transport behavior: Maxwell-Boltzmann distribution ($\alfa=2$)}
The Maxwell-Boltzmann distribution, 
\be
p_{\bs\eta}{\ss(\bs\eta)}  = \frac{1}{(2\pi)^{3/2}\gamma^3} \exp\(- \frac{|\bs\eta-\bs{\bar{\eta}}|^2}{2\gamma^2}\) ~,
\label{eq_MB_transport1}
\ee
corresponds to $\alfa = 2$ and $q = \nu$ (Table \ref{tab_similarity}).
Due to the exponential decay of \eqref{eq_MB_transport1}, only small displacements will be important in \eqref{eq_1D_dcdt_1}.  Thus, we expand in Taylor series (switching to index notation for clarity)
\be
c(t, x_i - \eta_i \sqrt{\nu \,\delta t}; \eta_i) = c(t, x_i; \eta_i) - \eta_j \sqrt{\nu \,\delta t} \frac{\p c}{\p x_j} + \half \eta_j\eta_k (\nu \,\delta t) \frac{\p^2c}{\p x_j \p x_k}  + {\cal O}(\delta t^\frac{3}{2})  ~.
\label{eq_MB_transport2}
\ee
Inserting \eqref{eq_MB_transport2} and \eqref{eq_MB_transport1} into \eqref{eq_1D_dcdt_1}, we have:
\be
\begin{split}
\frac{\p \bar{c}}{\pt}  
= \lim\limits_{\delta t \ra 0} \iiint\limits_{\!\!\!-\infty}^{~~~\infty}
&  \frac{ \frac{\p c}{\p x_j} (-\eta_j \sqrt{\nu \,\delta t}) + \half \eta_j\eta_k (\nu \,\delta t) \frac{\p^2c}{\p x_j \p x_k} + {\cal O}(\delta t^\frac{3}{2}) }{\delta t} \, 
  \\[-0.5 em]
&\quad\quad\quad\quad\quad\quad  \cdot \frac{1}{(2\pi)^{3/2}\gamma^3} \exp\(- \frac{|\bs\eta-\bs{\bar{\eta}}|^2}{2\gamma^2}\) \, d\bs\eta  ~.
\end{split}
\label{eq_MB_transport3}
\ee
The integrals evaluate as follows
{\footnotesize
\begin{align}
\iiint\limits_{\!\!\!-\infty}^{~~~\infty} \frac{c \, \eta_j}{(2\pi)^{3/2}\gamma^3} \exp\(- \frac{|\bs\eta-\bs{\bar{\eta}}|^2}{2\gamma^2}\) \, d\bs\eta &= \bar{c} \, \bar{\eta}_j  
\\ 
\iiint\limits_{\!\!\!-\infty}^{~~~\infty} \frac{c \, \eta_j \eta_k}{(2\pi)^{3/2}\gamma^3} \exp\(- \frac{|\bs\eta-\bs{\bar{\eta}}|^2}{2\gamma^2}\) \, d\bs\eta &= \bar{c} \, (\bar{\eta}_j\bar{\eta}_k + \gamma^2 \delta_{jk}) ~.
\end{align}
}
Thus, \eqref{eq_MB_transport3} reduces to
\be
\frac{\p \bar{c}}{\pt}  
= \lim\limits_{\delta t \ra 0}  \[   - \frac{\p\bar{c}}{\p x_j} \frac{\bar{\eta}_j \sqrt{\nu \delta t}}{\delta t}  + \half\frac{\p^2\bar{c}}{\p x_j \p x_k} \nu (\bar{\eta}_j\bar{\eta}_k +  \gamma^2 \delta_{jk}) + {\cal O}(\delta t^\frac{1}{2})  \]   ~.
\ee
Note that the mean displacement is equivalently given by $\bbxi = \bs{\bar{\eta}} \sqrt{\nu \delta t} = \ub \, \delta t$, where $\ub$ is the mean velocity.  Thus, for fixed mean velocity, 
$
\lim\limits_{\delta t \ra 0} \frac{\bar{\eta}_j \sqrt{\nu \delta t}}{\delta t}= \lim\limits_{\delta t \ra 0} \frac{\bar{u}_j \delta t}{\delta t} = \bar{u}_j
$
, while
$
 \lim\limits_{\delta t \ra 0} \bar{\eta}_j = \lim\limits_{\delta t \ra 0} \frac{\bar{u}_j \delta t}{ \sqrt{\nu \delta t}} = 0
$.  
Since $\gamma$ is a free parameter, we can choose $\gamma = \sqrt{2}$ so that we can interpret $\nu$ as the diffusivity.  Then, we achieve at the final result 
\begin{equation}
\frac{\p \bar{c}}{\pt}  + \ub \cdot \nabla\bar{c}  = \nu \nabla^2\bar{c}    ~,
\label{eq_1D_advection_diffusion_ensemble_average}
\end{equation}
which is the standard advection-diffusion equation.  Equation \eqref{eq_1D_advection_diffusion_ensemble_average} represents the ensemble-averaged behavior of \eqref{eq_3D_governing_eqn} when the probability distribution is Maxwell-Boltzmann.   It echoes the celebrated result of \citet{Einstein1905}, who showed that a random walk following the Maxwell-Boltzmann distribution is equivalent to Fickian diffusion.

%------------------------------------------------
\subsubsection{Ensemble-averaged transport behavior: Cauchy distribution ($\alfa=1$)}
\label{sec_Cauchy_transport}
The Cauchy distribution, 
\be
p_{\bs{\tilde\eta}}{\ss(\bs{\tilde\eta})}  =  \frac{1}{\pi^2 \gamma^3} \frac{1 }{\( |\bs{\tilde\eta}/\gamma|^2 + 1  \)^2} 
\label{eq_C_transport1}
\ee
corresponds to $\alfa = 1$ and $q = u_*$ (Table \ref{tab_similarity}).
Due to the algebraic decay of \eqref{eq_C_transport1}, both small and large displacements will be important in \eqref{eq_1D_dcdt_1}.   Therefore, we find it convenient to recast the transport equation \eqref{eq_1D_dcdt_0} in displacement form:
\begin{equation}
\frac{\p \bar{c}}{\pt} 
= \lim\limits_{\delta t \ra 0} \iiint\limits_{\!\!\!-\infty}^{~~~\infty} ~  \frac{ c{\ss(t,\x-\bxi; \bxi)} - c{\ss(t,\x; \bxi)} }{\delta t}  \, p_{\bxi}{\ss(\bxi; \delta t)} \, d\bxi ~.
\label{eq_1D_dcdt_1_1}
\end{equation}
To proceed, define the total displacement as the average plus a deviation $\bxi = \bs{\bar{\xi}} + \bs{\tilde{\xi}}$, such that $d\bxi =  d\bs{\tilde{\xi}}$.  Since $\bs{\bar{\xi}} = \ub \,\delta t$ and $\ub$ is finite, then $\bs{\bar{\xi}}$ is small in the limit of $\delta t$ tending to zero. Therefore, we may expand $c(t,\x-\bxi; \bxi)$ in a Taylor series in $\bs{\bar{\xi}}$:
\be
c{\ss(t,\x-\bxi; \bxi)} = c{\ss(t,\x-\bs{\tilde{\xi}} - \bs{\bar{\xi}}; \bxi)}
=
c{\ss(t,\x-\bs{\tilde{\xi}}; \bxi)} - \bs{\bar{\xi}} \cdot \nabla c{\ss(t,\x-\bs{\tilde{\xi}}; \bxi)} + {\cal O}(\bs{\bar{\xi}}^2)  ~.
\label{eq_c_taylor}
\ee
Then \eqref{eq_1D_dcdt_1_1} becomes
\be
\begin{split}
\frac{\p \bar{c}}{\pt} 
= \lim\limits_{\delta t \ra 0} \iiint\limits_{\!\!\!-\infty}^{~~~\infty} ~ & \frac{ \big[ c{\ss(t,\x-\bs{\tilde{\xi}}; \bxi)} - \bs{\bar{\xi}} \cdot \nabla c{\ss(t,\x-\bs{\tilde{\xi}}; \bxi)} + {\cal O}((\ub \,\delta t)^2) \big] - c{\ss(t,\x; \bxi)} }{\delta t}  \,   \, p_{\btxi}{\ss(\btxi; \delta t)} \, d\btxi \, . 
\end{split}
\label{eq_1D_dcdt_1_2}
\ee

Equation \eqref{eq_G2g} prescribes 
\be
\begin{split}
p_{\btxi}{\ss(\btxi ; \delta t)} &= \frac{ p_{\bteta}{\ss(\bteta\,=\,\btxi /(u_* \, \delta t))} }{ (u_* \, \delta t)^3 } \\
&= \frac{1}{ (\gamma u_* \, \delta t)^3 }  \frac{1}{\pi^2} \frac{1}{\(\frac{|\btxi|^2}{(\gamma u_* \, \delta t)^2} + 1 \)^2}   
=   \frac{1}{\pi^2} \frac{\gamma u_* \,\delta t}{\( |\bs{\tilde{\xi}}|^2 + (\gamma u_* \,\delta t)^2  \)^2} ~.
\label{eq_Cauch_pxi}
\end{split}
\ee
Inserting \eqref{eq_Cauch_pxi} into \eqref{eq_1D_dcdt_1_2} yields
{\footnotesize
\be
\frac{\p \bar{c}}{\pt} 
=
\frac{\gamma u_*}{\pi^2} 
\dashint\!\!\!\dashint\!\!\!\dashint_{\!-\infty}^\infty
\frac{  c{\ss(t,\x-\bs{\tilde{\xi}}; \bxi)} - c{\ss(t,\x; \bxi)} }{ |\bs{\tilde{\xi}}|^4 }  d\bs{\tilde{\xi}}
-
\lim\limits_{\delta t \ra 0} \iiint\limits_{\!\!\!-\infty}^{~~~\infty} ~   \ub \,  \cdot \nabla c{\ss(t,\x-\bs{\tilde{\xi}}; \bxi)}  \, \, 
 \frac{1}{\pi^2} \frac{\gamma u_* \delta t}{\( |\bs{\tilde{\xi}}|^2 + (\gamma u_* \,\delta t)^2  \)^2} 
\, \, d\bs{\tilde{\xi}}~.
\label{eq_1D_dcdt_1_3}
\ee
}
In the gradient term, the limit $\delta t \ra 0$ makes the fraction in the integrand  behave as a delta function, picking off the value $\ub \cdot \nabla c{\ss(t,\x;\bxi)}$.  To see this, realize that as $\delta t \ra 0$, the integrand becomes zero for all $\bs{\tilde{\xi}} \neq 0$.  Therefore, $\bs{\tilde{\xi}}$ can be assumed to be small, and a Taylor series used 
$\nabla c{\ss(t,\x-\bs{\tilde{\xi}}; \bxi)} = \nabla c{\ss(t,\x; \bxi)} - \bs{\tilde{\xi}} \cdot \nabla(\nabla  c{\ss(t,\x; \bxi)}) + \O(|\bs{\tilde{\xi}}|^2)$.  The $\bs{\tilde{\xi}}$ term in this Taylor series integrates to zero by symmetry.  The remaining integral evaluates to unity:
\be
 \iiint\limits_{\!\!\!-\infty}^{~~~\infty} ~ 
 \frac{1}{\pi^2} \frac{\gamma u_* \delta t}{\( |\bs{\tilde{\xi}}|^2 + (\gamma u_* \,\delta t)^2  \)^2} 
\, \, d\bs{\tilde{\xi}}
=
1 ~,
\ee
and all that remains is $\ub \,  \cdot \nabla c{\ss(t,\x; \bxi)}$.  Thus, the transport equation reduces to
\be
\frac{\p \bar{c}}{\pt} + \ub \,  \cdot \nabla c{\ss(t,\x;\bxi)}
=
\frac{\gamma u_*}{\pi^2} 
\dashint\!\!\!\dashint\!\!\!\dashint_{\!-\infty}^\infty
\frac{  c{\ss(t,\x-\bs{\tilde{\xi}};\bxi)} - c{\ss(t,\x;\bxi)} }{ |\bs{\tilde{\xi}}|^4 }  d\bs{\tilde{\xi}} ~.
\ee
We can now take an ensemble average of both sides using \eqref{eq_c_avg} in displacement form:
\be
\bc{\ss(t,\x)} = \mediumiiint_{\!\!\!-\infty}^{\infty} ~ c{\ss(t,\x; \bxi)} \, p_{\bxi}{\ss(\bxi; \delta t)} \, d\bxi ~,
\label{eq_c_avg_disp}
\ee
and we can make the substitution $\x' = \x - \btxi$ to obtain the final result
\be
\boxed{
\frac{\p \bar{c}{\ss(t,\x)}}{\pt} + \ub \,  \cdot \nabla \bar{c}{\ss(t,\x)}
=
\frac{\gamma u_*}{\pi^2} 
\dashint\!\!\!\dashint\!\!\!\dashint_{\!-\infty}^\infty
\frac{  \bar{c}{\ss(t,\x')} - \bar{c}{\ss(t,\x)} }{ |\x' - \x|^4 }  d\x' }~.
\label{eq_dcdt_Cauchy_final}
\ee

Equation \eqref{eq_dcdt_Cauchy_final} predicts turbulent dispersion according to a fractional Laplacian.  
This is analogous to the fractional Laplacian appearing in the momentum equation (in \S\ref{sec_derivation}).

\subsubsection{Ensemble-averaged transport behavior: Any $0 < \alpha < 2$}

We now generalize Equation \eqref{eq_dcdt_Cauchy_final} to any value of $\alpha$ in the range $0 < \alpha < 2$.  Recall the development in \S\ref{sec_Cauchy_transport} that lead to Equation \eqref{eq_1D_dcdt_1_2}, which is rearranged here
\be
\frac{\p \bar{c}}{\pt} 
= \lim\limits_{\delta t \ra 0} \iiint\limits_{\!\!\!-\infty}^{~~~\infty} ~  \frac{ c{\ss(t,\x-\bs{\tilde{\xi}}; \bxi)} -  c{\ss(t,\x; \bxi)} }{\delta t}  \,   \, p_{\btxi}{\ss(\btxi; \delta t)} \, d\btxi 
-
\lim\limits_{\delta t \ra 0} \iiint\limits_{\!\!\!-\infty}^{~~~\infty} ~ 
\ub \cdot \nabla c{\ss(t,\x-\bs{\tilde{\xi}}; \bxi)} \, p_{\btxi}{\ss(\btxi; \delta t)}
 \, d\btxi \, . 
\label{eq_EAany_1}
\ee
where
\be
p_{\btxi}{\ss(\btxi ; \delta t)} = \frac{ p_{\bs{\tilde\eta}}{\ss(\bteta \,=\, \btxi   /(q \, \delta t)^{1/\alfa})} }{ (q \, \delta t)^{3/\alfa} } ~,
\ee
and $\btxi = \bxi - \bbxi$.
As in \S\ref{sec_Cauchy_transport}, the limit $\delta t \ra 0$ makes $\btxi /(q \, \delta t)^{1/\alfa}$ very large for any finite $\btxi$, which enables simplification of both terms in \eqref{eq_EAany_1}.  For the second term of \eqref{eq_EAany_1}, we note that as in \S\ref{sec_Cauchy_transport}, $p_{\btxi}{\ss(\btxi ; \delta t)}$ behaves as a delta function, so the second term simplifies to $\ub \,  \cdot \nabla c{\ss(t,\x; \bxi)}$.
For the first term of \eqref{eq_EAany_1}, we may limit 
ourselves to retaining only the tails of $p_{\btxi}{\ss(\btxi ; \delta t)}$, which at 3D, for an arbitrary value of $\alpha$ and with elasticity $\gamma$ 
are given by \citep{Nolan2006}
\begin{equation}
p_{\btxi}{\ss(\btxi ; \delta t)} 
 \simeq 
\frac{1}{(q \, \delta t)^{3/\alfa}}
\frac{\bar{C}_\alpha}{\gamma^3 ~| \btxi   /\gamma (q \, \delta t)^{1/\alfa}|^{\alpha+3}} 
= 
\frac{\gamma^\alpha   q\delta t \, \bar{C}_\alpha}{|\btxi|^{\alpha+3}}
\end{equation} 
with $\bar{C}_\alpha$ given later in \eqref{eq_Levy_tails}.
The first term in \eqref{eq_EAany_1} can thus be expressed as
{\small
\be
\begin{split}
\gamma^\alpha q \bar{C}_\alpha 
\dashint\!\!\!\dashint\!\!\!\dashint_{\!-\infty}^\infty
  \frac{ c{\ss(t,\x-\bs{\tilde{\xi}}; \bxi)} -  c{\ss(t,\x; \bxi)} }{|\btxi|^{\alpha+3}}   
    \lim_{\delta t \rightarrow 0} \left\{ 
    \frac{\delta t}{\delta t} 
    \right\}
      d\btxi 
=
\gamma^\alpha q \bar{C}_\alpha
\dashint\!\!\!\dashint\!\!\!\dashint_{\!-\infty}^\infty
   \frac{ c{\ss(t,\x-\bs{\tilde{\xi}}; \bxi)} -  c{\ss(t,\x; \bxi)} }{|\btxi|^{\alpha+3}}  
      d\btxi ~.
\end{split}
\label{eq_dcdt_3}
\ee
}
This straightforward and non-trivial limit 
is the result of the stability property of the distribution $p_{\bs{\tilde\eta}}{\ss(\bs{\tilde\eta})}$, and 
this verifies the necessity of a stable distribution.

Upon inserting \eqref{eq_dcdt_3} into \eqref{eq_EAany_1}, taking an ensemble average of both sides using \eqref{eq_c_avg_disp}, and making the substitution $\x' = \x - \btxi$, we obtain the final result
\be
\boxed{
\frac{\p \bar{c}{\ss(t,\x)}}{\pt} + \ub \,  \cdot \nabla \bar{c}{\ss(t,\x)} =  \gamma^\alpha q \bar{C}_\alpha  
\dashint\!\!\!\dashint\!\!\!\dashint_{\!-\infty}^\infty
  \frac{ \bar{c}{\ss(t,\x')} -  \bar{c}{\ss(t,\x)} }{|\x'-\x|^{\alpha+3}}  
      d\x' 
}~.
\label{eq_transport_general_final}
\ee
Equation \eqref{eq_transport_general_final} predicts turbulent dispersion according to a fractional Laplacian, as a generalization of \eqref{eq_dcdt_Cauchy_final} to fractional order $\alfa$.

In this section, we have shown that the canonical diffusion operator $\nu \nabla^2 \bar{c}$ is uniquely tied to similarity variable $\bs\eta = \bxi/\sqrt{\nu \,\delta t}$, which has an ${\cal O}(\sqrt{\nu \delta t})$ scaling on the diffusion distance.  Similarly, we showed that the case $\alfa = 1$ is uniquely tied to similarity variable $\bs\eta = \bxi/(u_* \delta t)$.  It is well known that in shear turbulent flows, mixing occurs over distances that scale by ${\cal O}(u_* \delta t)$, where $u_*$ is some appropriate characteristic eddy velocity (such as the friction velocity) \citep{CushmanRoisin2006, Kampf2016}.  Thus, the similarity parameter $\bs\eta = \bxi / (u_* \, \delta t)$ and Cauchy distribution are appropriate for modeling shear turbulent flows.  

We also showed that the $\delta t \ra 0$ limit demands the velocity distribution be {\it stable}.  Further, we showed that the L\'evy $\alfa$-stable distributions satisfy this stability constraint and thus provide a family of distributions with which to model laminar through turbulent flows.

%----------------------------------------------------------------------------------------------------
%----------------------------------------------------------------------------------------------------
%----------------------------------------------------------------------------------------------------
%----------------------------------------------------------------------------------------------------
%----------------------------------------------------------------------------------------------------
%----------------------------------------------------------------------------------------------------
%----------------------------------------------------------------------------------------------------
%----------------------------------------------------------------------------------------------------
%----------------------------------------------------------------------------------------------------
%----------------------------------------------------------------------------------------------------
%----------------------------------------------------------------------------------------------------
%----------------------------------------------------------------------------------------------------
\section{Derivation of the Fractional Laplacian as a Turbulence Model}
\label{sec_derivation}
\subsection{Boltzmann Kinetics}
\label{sec_Boltzmann}
%-------------------------------------------------------------------------------------------
Our premise is {\it Boltzmann kinetic theory}, wherein the flowfield is described by
the {\it mass probability density function} $f{\ss(t,\x,\u)}$ with time $t$, particle position $\x=[x,y,z]$, and particle velocity $\u=[u,v,w]$ as independent variables. 
By definition, 
$
f(t,\x,\u) \, d\x \, d\u
$
is the mass 
of fluid particles at time $t$
located within volume $d\x$  
surrounding position $\x$    
that have velocities within the range $d\u$  
surrounding velocity $\u$.    
The ensemble-averaged {\it hydrodynamic quantities} are derived from $f$ via integrals over all possible velocities \citep{Chen2011}:
\begin{align}
\text{density} && \rho{\ss(t,\x)} &= \mediumiiint_{\!\!-\infty}^\infty  ~    f{\ss(t,\x,\u)} ~d\u  
\label{eq_hydro_density}  \\
\text{velocity} &&   \bar{u}_i{\ss(t,\x)} &= \tfrac{1}{\rho}\mediumiiint_{\!\!-\infty}^\infty  ~ u_i ~f{\ss(t,\x,\u)} ~d\u   
\label{eq_hydro_velocity}  \\
\text{specific internal energy} &&  \check{u}{\ss(t,\x)}  = \tfrac{3}{2} U^2 &= \tfrac{1}{\rho} \mediumiiint_{\!\!-\infty}^\infty  ~ \half |\u-\ub{\ss(t,\x)}|^2 ~f{\ss(t,\x,\u)} ~d\u 
\label{eq_hydro_energy}
\end{align}
where $\bar{u}_i$ is the $i^{th}$ component of the Eulerian flow velocity, and 
$U$ is the thermal agitation speed, which may be understood as the speed of a particle between collisions.  
Note that the mass distribution $f{\ss(t,\x,\u)}$ is a function of seven independent variables (time, 3D space, and 3D velocity space).
\footnote{Using the fact that $|\u - \ub|^2 = |\u|^2 - 2\ub\cdot\u + |\ub|^2$, it is easy to show that the {\it specific total energy} $\check{e} = \tfrac{1}{\rho} \iiint \half |{\bf v}|^2  f \, d{\bf v} = \check{u} + \half|\ub|^2$ is the sum of the specific internal energy  $\check{u}$ and the specific kinetic energy $\half|\ub|^2$.}
\footnote{Unless otherwise noted, we use bold variables to indicate vectors and  use Einstein's convention of summing over repeated indices.  Also, we will hereafter use the shorthand $\iiint (\dots\!) \, d\u$ to imply the definite integral $\iiint_{-\infty}^\infty (\dots\!) \, du \, dv \, dw$. }

The evolution of the mass distribution $f{\ss(t,\x,\u)}$ is governed by the {\it Boltzmann equation}, with the classical BGK \citep{Bhatnagar1954} formulation for the collision term on the right hand side  \citep{Succi2001}:
\begin{equation}
       \tfrac{\partial f}{\partial t} 
+ \u \cdot \nabla f
= \tfrac{1}{\tau} (f^{eq}_\alfa - f) \, ,
\label{eq_Boltzmann}
\end{equation}
where $f^{eq}_\alfa{\ss(t,\x,\u)}$ is an {\it equilibrium distribution} to which $f$ relaxes, and $\tau$ is the {\it relaxation time}, which can be understood as the time between successive collisions of a particle \citep{Succi2001}.  
 The equilibrium distribution 
$f_\alpha^{eq}$ has a prescribed structure and must share certain moments with 
the actual distribution $f$: 
The free parameters defining $f^{eq}_\alfa$ must be such that the hydrodynamic variables  ($\rho,\ub,\check{u}$) are recovered when inserting $f^{eq}_\alfa$ into \eqref{eq_hydro_density}--\eqref{eq_hydro_energy}, since $f^{eq}_\alfa$ is a special case of $f$.  This requirement ensures that the collision term (right hand side) in the Boltzmann Equation \eqref{eq_Boltzmann} conserves mass, momentum and energy.

The equilibrium distribution $f_\alpha^{eq}$ may be taken as the traditional Maxwell-Boltzmann 
distribution or any other member of the family of the L\'evy $\alfa$-stable distributions, as shown in \S\ref{sec_transport}.  The only condition needed for now is that $f_\alpha^{eq}$ be 
{\it isotropic} with respect to its velocity variables, and thus of the form:
\begin{equation}
f^{eq}_\alfa{\ss(t,\x,\u)} \equiv \frac{\rho}{U^3} F(\Delta{\ss(t,\x,\u)}) 
\label{eq_feq_F}
\end{equation}
in which $\Delta{\ss(t,\x,\u)}$ is short-hand for
\begin{equation}
\Delta{\ss(t,\x,\u)} \equiv  \frac{|\u - \ub{\ss(t,\x)}|^2}{U^2}  \, .
\label{eq_Delta}
\end{equation}
Note that the distribution is multivariate in three dimensions.  
Two important examples are as follows:
\begin{align}
\text{Maxwell-Boltzmann ($\alfa=2$):} 	\quad  F(\Delta) &=  \frac{1}{(2\pi)^{3/2}} e^{-\Delta/2} ~;
 \label{eq_F_MB} \\	 
\text{Cauchy ($\alfa=1$):} 				\quad  F(\Delta) &=  \frac{1}{\pi^2}  \frac{1}{(\Delta + 1)^2}  ~.
  \label{eq_F_C}   
\end{align}

A concern arises whether higher moments are finite when the equilibrium function decays 
algebraically instead of exponentially.  For the Cauchy distribution, for example, 
the integrals for the second and higher moments diverge, raising objections about the value 
of the ensuing analysis.  In any physical situations, however, the spatial domain is 
always finite, and fluctuating velocities remain within bound.  Thus, no integration  
ever needs to be carried to infinity, and all moments remain finite.  For the purpose 
of clarity and simplicity in the presentation, we write our integrals as if they could 
extend to infinity but do imply that their domain of integration remains finite, however 
extended it might be.  For those integrals for which the domain can be unlimited without 
causing mathematical difficulties, we do carry the integrations to infinity, but only 
for pure mathematical convenience fully realizing that infinity is only a simplifying 
mathematical construct and that it can never be so in an actual situation.

The following analysis is performed without further specification of $f_\alpha^{eq}$ 
until needed at a later stage.  Note, however, that $f^{eq}_\alfa$ is an even function of each $(u_i - \bar{u}_i)$, so integrals over odd powers of $(u_i - \bar{u}_i)$ integrate to zero, which will simplify the subsequent analysis.

\subsection{Hydrodynamic Equations}

The ensemble-averaged {\it hydrodynamic equations} for mass and momentum
 are recovered by multiplying the Boltzmann equation by 1 or $\u$ 
 and integrating over the velocity space. 
 The only ``trick'' is that $\u$ is an independent variable, so it can ``hop into'' the derivatives, and the order of the derivatives and velocity integrations can be switched.  For example, 
$\iiint (\u \cdot \nabla f) \, d\u = 
\nabla\cdot (\iiint  \u f \, d\u) = 
 \nabla\cdot(\rho\ub)$.  
 In this way, $\iiint (\textit{\small Boltzmann}) \, d\u$ yields the {\it continuity equation}
 \begin{equation}
\frac{\partial \rho}{\partial t} + \nabla\cdot(\rho\ub) = 0 ~,
\label{eq_mass}
\end{equation}
and 
$\iiint \u \, (\textit{\small Boltzmann}) \, d\u$ yields the momentum equation as follows:
\begin{equation}
\mediumiiint \big\{    
 \tfrac{\p(\u f)}{\pt}  +  \nabla \cdot ({\bf uu}^\T  f)
\big\} \,d\u
=  \mediumiiint \u \tfrac{1}{\tau} (f - f_\alfa^{eq}) \,d\u  ~.
\label{eq_momentum_conservative_integral}
\end{equation}
The right hand side is zero, since both $f$ and $f_\alfa^{eq}$ obey \eqref{eq_hydro_density}.
The unsteady term evaluates to $\tfrac{\p(\rho\ub)}{\pt}$ by virtue of \eqref{eq_hydro_velocity}.  
The advective term can be evaluated by first noting that $(\u-\ub)(\u-\ub)^\T = \u\u^\T - \u\ub^\T - \ub\u^\T + \ub\ub^\T$.  Then,    
\be
\mediumiiint  \nabla \cdot ({\bf uu}^\T  f) \,d\u 
=
\nabla \cdot \Big[ \underbrace{  \mediumiiint  (\u-\ub)(\u-\ub)^\T  f \,d\u }_{\equiv -  \bs{\sigma}}
+
 \underbrace{ \mediumiiint  (\u\ub^\T + \ub\u^\T -  \ub\ub^\T)  f \,d\u}_{=\rho \ub\ub^\T}  \Big] ~.
\ee
Thus, \eqref{eq_momentum_conservative_integral} becomes 
$
  \tfrac{\partial(\rho \ub)}{\partial t  }
+ \nabla\cdot(\rho \ub \ub^\T)
= \nabla\cdot\bs{\sigma},
$
which can be simplified using \eqref{eq_mass} and the definition
$\frac{D\ub}{Dt} \equiv  \frac{\p\ub}{\pt} + \ub\cdot\nabla\ub$ 
to yield the ensemble-averaged {\it momentum equation}:
\begin{equation}
\rho \frac{D\ub}{Dt}
= \nabla\cdot\bs{\sigma}
\label{eq_momentum}
\end{equation}
in which the {\it ensemble-averaged stress tensor} $\bs\sigma$ is defined as
\begin{equation}
\boxed{
\sigma_{ij}{\ss(t,\x)} \equiv - \mediumiiint (u_i - \bar{u}_i{\ss(t,\x)})(u_j - \bar{u}_j{\ss(t,\x)}) \, f{\ss(t,\x,\u)} \,d\u 
} \, . 
\label{eq_sigma_def}
\end{equation}
Note the similarity between \eqref{eq_sigma_def} and  a {\it Reynolds stress}, with the velocity departures from their respective means playing 
the role of Reynolds' velocity fluctuations and the integral over $f d\u$ playing the role of Reynolds' ensemble average.  
However, the stress \eqref{eq_sigma_def} is {\it not} a Reynolds stress:
\begin{itemize} 
\item Expression \eqref{eq_sigma_def} is a definition within the framework of kinetic theory;  and 
\item Expression \eqref{eq_sigma_def} is rooted in the Boltzmann Equation (rather than Navier-Stokes).
\end{itemize} 

It is widely accepted that turbulence arises from the nonlinear advective term in the Navier-Stokes equations, $\v\cdot\nabla\v$.  Indeed, with a Reynolds decomposition $\v = \vb + \vt$, the Reynolds stresses arise as follows $\overline{\v\cdot\nabla\v}= \overline{(\vb+\vt)\cdot\nabla(\vb+\vt)} = \vb\cdot\nabla\vb + \overline{\vt\cdot\nabla\vt}$.  Within the kinetic theory framework, both the advection term $\ub\cdot\nabla\ub$ and the force term $\nabla\cdot\bs{\sigma}$ arise from the advection term in the Boltzmann equation, $\u \cdot \nabla f$.  Since the Boltzmann equation is the parent of the Navier-Stokes equations, and since the turbulent force arises from the advection term as desired, we expect that the force $\nabla\cdot\bs{\sigma}$ is a reasonable alternative to the Reynolds force $\overline{\vt\cdot\nabla\vt}$.

Moving on, our goal is to obtain closed-form expressions for the stress $\bs{\sigma}$ and force $\nabla\cdot\bs{\sigma}$ that depend only on the ensemble-averaged macroscopic quantities $\rho$, $\ub$, and $q$.
We seek to evaluate the stress tensor  
for the Maxwell-Boltzmann, Cauchy, and general L\'evy $\alfa$-stable distributions.
If $f^{eq}_\alfa$ falls off rapidly (with $\u$), as is the case when $f^{eq}_\alfa$ is {\it Maxwell-Boltzmann} ($\alfa=2$), then large velocity deviations are very unlikely, and the stress is viscous.  
If $f^{eq}_\alfa$ has ``heavy tails'', as is the case for  other L\'evy $\alfa$-stable $f^{eq}_\alfa$ ($\alfa<2$), then large velocity fluctuations are permitted, as in turbulence.

Note that in traditional studies of Boltzmann kinetics, the relaxation time $\tau$ is assumed to be 
short and the equilibrium distribution $f^{eq}_\alfa$ assumed to be Maxwell-Boltzmann.  With these assumptions, the {\it Chapman-Enskog perturbation expansion} can be used to recover the Navier-Stokes equations, with constitutive relations consistent with those for a viscous conductive ideal gas \citep{Chapman1991}. 
Herein, we pursue a different approach.  We first determine the analytic solution the Boltzmann equation, expressing $f$ in terms of $f_\alfa^{eq}$ in \eqref{eq_f_general_soln} below, and then we use this $f$ to evaluate the stress, as defined in \eqref{eq_sigma_def}.

\subsection{Mass Probability Distribution}
\label{sec_feq}
Equation \eqref{eq_Boltzmann} is linear in its unknown variable $f{\ss(t,\x,\u)}$ and possesses the 
following analytical solution for the {\it mass probability distribution:}
\begin{equation}
f{\ss(t,\x,\u)} ~=~ \tfrac{1}{\tau} \mediumint_{-\infty}^{~t} ~
f^{eq}_\alfa{\ss(t',\x - (t-t')\u,\u)} ~e^{-\frac{t-t'}{\tau}} ~dt' \, .
\label{eq_f_general_soln}
\end{equation}
Equation \eqref{eq_f_general_soln} prescribes $f{\ss(t,\x,\u)}$ as the weighted sum of the particles with velocity $\u$, with the weight being an exponential attenuation to account for the scattering of particles due to collisions between their earlier location, $\x' \equiv \x - (t-t') \u$, and their current location, $\x$.  

To simplify the algebra, we find it convenient to define the
dimensionless flight time $s \equiv (t-t')/\tau$ such that  $ds = -dt'/ \tau$, $\x' = \x - s\tau\u$, and
$d\x' = (dx')(dy')(dz') = (-s\tau \, du) (-s\tau \, dv) (-s\tau \, dw) = (-s\tau)^3 \, d\u$.
With this notation, \eqref{eq_f_general_soln} takes the more compact form:
\begin{equation}
\boxed{
f{\ss(t,\x,\u)} ~=~   \mediumint_0^\infty f^{eq}_\alfa{\ss(t-s\tau,\x - s\tau\u,\u)} ~e^{-s} ~ ds
} \, .
\label{eq_f_general_soln_2}
\end{equation}
Unless otherwise noted, we will hereafter use the shorthand $\int(\ldots)ds$ in lieu of  $\int_0^\infty(\ldots)ds$.

\subsection{Scales, Definitions, and Assumptions}
\label{sec_scales}
\noindent
We define the following quantities:
\begin{itemize}

\item[$L$] characteristic length scale over which changes occur in the macro-scale hydrodynamics.  Examples include the boundary layer thickness, pipe diameter, or channel width. 

\item[$V$] characteristic velocity scale of the flow field, which may be taken as the value ascribed at a boundary, the freestream speed, or a suitable temporal/spatial average;

\item[$U$] thermal agitation speed of the particles, appearing in the definition of $f^{eq}_\alfa$ \eqref{eq_feq_F};

\item[$\tau$] relaxation time (time between particle collisions) appearing in \eqref{eq_Boltzmann}; 

\item[$\ell$] demarcation length scale that separates ``small'' from ``large'' particle displacements;

\item[$\lambda$] {\it mean free path}, which is the average distance covered by a particle between successive collisions:
\begin{equation}
\lambda = U \tau  \, .
\label{eq_mean_free_path}
\end{equation} 

\end{itemize}

\noindent
The {\it Knudsen number} is defined as the ratio of the mean free path to the length scale of the flow:
\begin{equation}
Kn = \frac{\lambda}{L} = \frac{U \tau}{L} ~.
\end{equation}
Using these quantities, we make the following assumptions:

~\\
\noindent {\bf Assumption \#1}:  Herein, we restrict our attention to incompressible, isothermal flows.  That is, we assume that the density $\rho$ and thermal agitation speed $U$ are both constant, from which follows $\tfrac{\p\bar{u}_i}{\p x_i} = 0$ (implied summation over $i=1,2,3$) from \eqref{eq_mass}.

~\\
\noindent {\bf Assumption \#2}:  We need only consider dimensionless time delays on the order of unity:
\begin{equation}
s = \mathcal{O}(1) .
\label{eq_s_vs_unity}
\end{equation}
The reason for this is readily apparent from examination of \eqref{eq_f_general_soln_2}.  
The $e^{-s}$ attenuation factor renders any value much larger than unity inconsequential.

~\\
\noindent {\bf Assumption \#3}:  Assume that the characteristic flow speed is less than or on the order of the thermal speed:
\be
V \leq \O(U) ~.
\label{eq_V_ll_U}
\ee
Moreover, since $|\ub| \sim \O(V)$ by definition, Equation \eqref{eq_V_ll_U} implies $|\ub| \leq \O(U)$.

~\\
\noindent {\bf Assumption \#4}:  We assume that the following ordering of scales exists:
\be
\lambda \ll \ell \ll L  ~.
\label{eq_L_scales}
\ee
One physically-meaningful definition of $\ell$ is the {\it Kolmogorov microscale}, which is the size of the smallest eddies of the turbulent cascade, where they get extinguished by molecular viscosity. Appendix \ref{sec_Kolmogorov} shows that this would be the case if $\ell$ were chosen according to 
\be
\frac{\ell}{L} = Re^{-3/4}  ~,
\label{eq_ell}
\ee
which is very much less than unity for a high-Reynolds-number turbulent flow.  This demarcation makes sense, because it places the cut between small and large displacements precisely where turbulence ends and viscosity takes over.  While Equation \eqref{eq_ell} justifies the second inequality in \eqref{eq_L_scales}, the first inequality $\lambda \ll \ell$ is merely the classical assumption of a continuum, even at the smallest eddy scale.  
Requiring that both inequalities hold implies $\lambda  \ll L$ ({\it i.e.} a very small Knudsen number), 
which is a classic assumption in the context of Boltzmann kinetics \citep{Succi2001}.
As a corollary to \eqref{eq_L_scales}, we demand that $Re\, Kn = V\lambda/\nu \ll 1$, which means that the Knudsen number must be much smaller than the Reynolds number is large;  this implies moderate speeds, consistent with the incompressibility assumption above.

This ordering of scales \eqref{eq_L_scales} will permit the analysis in sections \S\ref{sec_small} and \S\ref{sec_large},  allowing us to consider the cases of ``small displacements'' ($s\tau|\u| < \ell$) and ``large displacements'' ($s\tau|\u| > \ell$) separately.  For small displacements, the assumption $\ell \ll L$ permits the use of a Taylor series expansion to simplify the analysis in \S\ref{sec_small}.  For large displacements, the assumption $\lambda \ll \ell$ has two important implications used in \S\ref{sec_large}: 

(1) large speeds, $|\u| \gg |\ub|$; and 

(2) large velocity deviations $\Delta \equiv |\u - \ub|^2/U^2 \gg 1$. 

 These implications are derived as follows: Since $s \sim \O(1)$ by virtue of \eqref{eq_s_vs_unity}, large displacements ($s\tau|\u| > \ell$) imply $|\u| > \O(\ell/\tau) = \O(\ell/\lambda) U \gg U$.  Since $|\ub| \leq \O(U)$ by virtue of \eqref{eq_V_ll_U}, $|\u| \gg U$ then implies $|\u| \gg |\ub|$.  Then, $\Delta \equiv |\u - \ub|^2/U^2 = |\u|^2/U^2 - 2 \u \cdot \ub/U^2 + |\ub|^2/U^2 \gg 1$.

~\\
\noindent {\bf Assumption \#5}:  We assume that the temporal evolution of the flow field proceeds 
on a time scale much longer than the relaxation time $\tau$:
\begin{equation}
\frac{\partial}{\partial t} << \frac{1}{\tau} ~.
\label{eq_time_vs_tau}
\end{equation}
This is justified by considering the ratio $V\tau/L$, which by virtue of \eqref{eq_V_ll_U} is less than $\O(U\tau/L)$. Now since $U\tau = \lambda \ll L$ by virtue of \eqref{eq_L_scales}, it is clear that $V\tau/L \ll 1$.  This implies $\tau \ll L/V$, \ie the relaxation time is much shorter than the time scale of the flowfield.  Phenomenologically, this assumption is consistent with the fact that the evolution of flow as a whole occurs as a result of a great many collisions among particles.  

This assumption permits us to perform Taylor expansions in time of functions with the time shift $t - s\tau$ (e.g. to simplify \eqref{eq_f_general_soln_2}).
This classical assumption of Boltzmann kinetics also justifies the use of the BGK collision model $\tfrac{1}{\tau} (f^{eq}_\alfa - f)$, which essentially is a finite difference approximation to the rate of change of $f$ that is brought about by particle collisions.

~\\
\noindent {\bf Assumption \#6}:  For reasons that were made clear in \S\ref{sec_transport}, we assume that any admissible $f_\alfa^{eq}$ satisfy two properties: 
\begin{itemize}

\item $f_\alfa^{eq}$ must be {\it isotropic} (equal probability in all velocity directions, such that $f^{eq}_\alfa$ depends only on the speed $|\u-\ub|$ and {\it not} on the direction $(\u-\ub)/|\u-\ub|$ of the molecular velocity); and 

\item $f_\alfa^{eq}$ must be {\it stable} (the sum of two independent and identically-distributed random variables has the same probability distribution as the random variables themselves).  

\end{itemize}

Notably, we do {\it not} assume that the velocity components $(u,v,w)$ are {\it independent, identically-distributed} random variables, which would imply
$f^{eq}{\ss(t,\x,\u)} = f^{eq}{\ss(t,\x,u)} f^{eq}{\ss(t,\x,v)} f^{eq}{\ss(t,\x,w)}$.
In developing statistical mechanics, Maxwell assumed both {\it isotropy} and {\it independence}, and the only function that satisfies both these requirements is the {\it normal distribution} \eqref{eq_F_MB} \citep{Chapman1991}.  For counterexample, the multivariate Cauchy distribution \eqref{eq_F_C} is isotropic and stable, but it is {\it not} the product of three independent univariate Cauchy distributions.

We argue that the velocity components $(u,v,w)$ are {\it not} independent, because the velocity magnitude is related to the temperature of the gas; therefore, for a given temperature (speed budget), one expects that a molecule with very large $u$ (consuming the entire speed budget) should have very small $v$ and $w$. 
 These three velocity components may be {\it uncorrelated}, as implied by \eqref{eq_feq_F}, without making the additional unnecessary restriction that the velocity components are {\it independent}.
In other words, we argue that Maxwell made an assumption that was unnecessary, and if we remove his assumption, then we are led to admit the entire family of {\it L\'evy $\alfa$-stable distributions} (of which the Maxwell-Boltzmann distribution is a special case $\alfa=2$).

\begin{figure}
\begin{center}
\includegraphics[width=\textwidth]{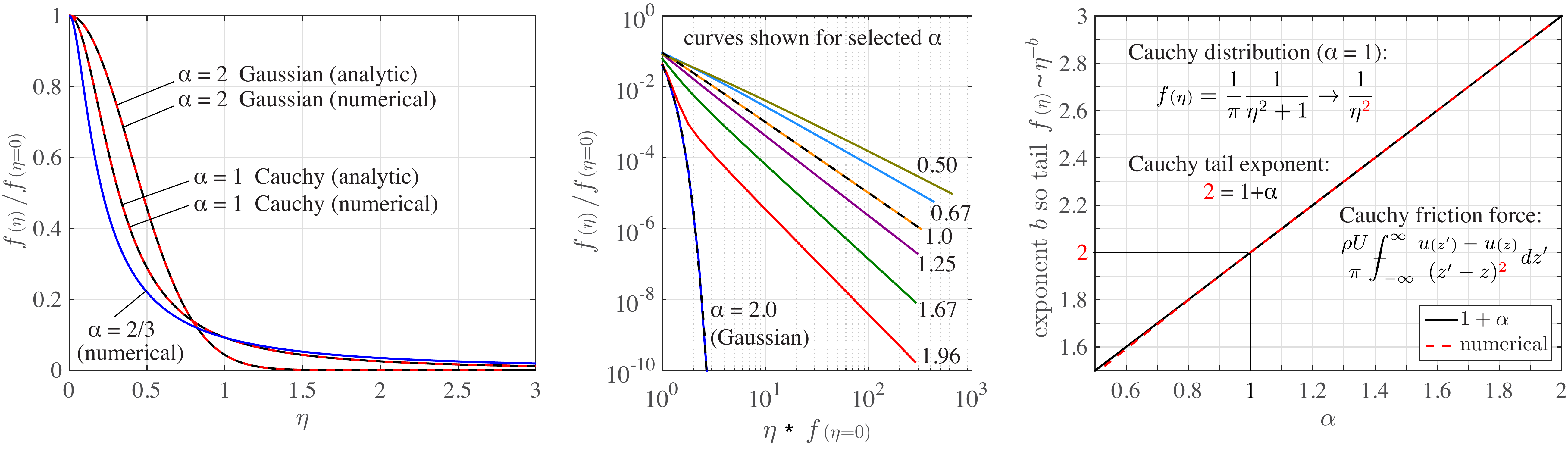}
\caption{[left] Univariate {\it L\'evy $\alfa$-stable probability distributions}, of which Maxwell-Boltzmann ($\alfa=2$) and Cauchy ($\alfa=1$) are special cases.  [center] Tails of these distributions for several $\alfa$. [right]  Log-slope of the tail of the distributions, comparing numerical results from the center figure with the theoretical asymptotic value $1+\alfa$.  These log-slopes are identical to the exponent of the fractional Laplacian (in one dimension). 
}
\label{fig_Levy}
\end{center}
\end{figure}

Table \ref{tab_pdf3D} lists the formulae for isotropic multivariate L\'evy $\alpha$-stable distributions.   Figure \ref{fig_Levy} shows the univariate distributions for $\alfa = 2, 1$, and 2/3, as well as the tails of the distributions for several $\alfa$.  In \S\ref{sec_tau_turb_Levy}, we will find that the exponent in the tail of the distribution is related to the exponent in the fractional Laplacian.

\begin{table}
\caption{Probability density functions $f{\ss(\u)}$ in 1, 2, 3, and $n$ dimensions for random variable $\u \in \mathbb{R}^n$ with ``mean'' (location parameter) $\ub$ and ``standard deviation'' (scale parameter) $U$.  Here, we assume the probability distributions are {\it isotropic}, so they are only a function of the square magnitude $\Delta \equiv |\bs\upsilon|^2$ of the standard normal variable $\bs\upsilon \equiv (\u - \ub)/U$. 
 In the general L\'evy $\alfa$-stable case, we use the notation $d{\bf k} = dk_1\,dk_2\ldots dk_n$. 
In the Cauchy case, we define $c \equiv (n+1)/2$. }\vspace{0.5 em}
\begin{center}
\small
\begin{tabular}{c @{\hskip 5mm} l @{\hskip 5mm} l @{\hskip 5mm} l}
{\it Dim.}	& {\it Maxwell-Boltzmann} & {\it Cauchy} & {\it general L\'evy $\alfa$-stable} \\[0.5 em]
1
&
$f{\ss(\upsilon)} =   \tfrac{1}{(2\pi)^{1/2}}~e^{ -\upsilon^2/2 } $
&
$f{\ss(\upsilon)} =  \tfrac{1}{\pi}    \tfrac{1}{\upsilon^2+1}$
&
$f{\ss(\upsilon)} = \tfrac{1}{ 2\pi}       \mediumint_{\!\!-\infty}^\infty \, e^{-|k|^\alfa-ik \upsilon} \,dk$
\\
2
&
$f{\ss(\bs\upsilon)} =  \tfrac{1}{2\pi}          ~ e^{ -|\bs\upsilon|^2/2 }$
&
$f{\ss(\bs\upsilon)} =  \tfrac{1}{2\pi}   \tfrac{1}{(|\bs\upsilon|^2+1)^{3/2} }$
&
$f{\ss(\bs\upsilon)} = \tfrac{1}{(2\pi)^2}   \mediumiint_{\!\!-\infty}^\infty   \,e^{-|{\bf k}|^\alfa -i {\bf k}\cdot\bs\upsilon} \,d{\bf k} $
\\
3
&
$f{\ss(\bs\upsilon)} =  \tfrac{1}{(2\pi)^{3/2}} ~  e^{ -|\bs\upsilon|^2/2 }$
&
$f{\ss(\bs\upsilon)} =  \tfrac{1}{\pi^2}  \tfrac{1}{(|\bs\upsilon|^2+1)^2 }$ 
&
$f{\ss(\bs\upsilon)} = \tfrac{1}{(2\pi)^3}  \mediumiiint_{\!\!-\infty}^\infty \,e^{-|{\bf k}|^\alfa -i {\bf k}\cdot\bs\upsilon } \,d{\bf k}  $
\\
$n$
&
$f{\ss(\bs\upsilon)} =  \tfrac{1}{(2\pi)^{n/2}} ~  e^{ -|\bs\upsilon|^2/2 } $
&
$f{\ss(\bs\upsilon)} =  \tfrac{\Gamma\(c\)}{\pi^c}  \tfrac{1}{(|\bs\upsilon|^2+1)^c }$
&
$f{\ss(\bs\upsilon)} = \tfrac{1}{(2\pi)^n}  \mediumint\!\!\dots\!\!\mediumint_{\mathbb{R}^n} \,e^{-|{\bf k}|^\alfa -i {\bf k}\cdot\bs\upsilon } \,d{\bf k}$
\end{tabular}
\normalsize
\end{center}
\label{tab_pdf3D}
\end{table}%

For example numbers, consider dry air at standard temperature $T = 15^\circ$C 
and pressure $p = 101.3$ kPa.  
At these conditions, laboratory measurements give the following values: 
{\it density}           $\rho = 1.23$ kg/m$^3$,     
{\it dynamic viscosity}  $\mu = 1.79 \times 10^{-5}$ kg/m.s, 
{\it kinematic viscosity} $\nu = \mu/\rho = 1.46 \times 10^{-5}$ m$^2$/s, and
{\it mean free path} $\lambda = 68 \times 10^{-9}$ m.   
The {\it thermal speed} is $U = \sqrt{3 k_B T/m} = 502$ m/s, where $k_B = 1.38 \times 10^{-23}$ J/K is Boltzmann's constant, and $m = (29$ g/mol) $/ (6.02 \times 10^{23}$ molecules/mol) $= 4.82\times 10^{-26}$ kg/particle is the molecular weight.  
Using these numbers, we can infer the {\it collision time} $\tau = \lambda/U = 1.35 \times 10^{-10}$ s.  
Equation \eqref{eq_tau_visc_final} predicts a laminar viscosity of 
$\mu = \rho U^2 \tau = 1.23$ kg/m$^3 * (502$ m/s$)^2 * 1.35 \times 10^{-10}$ s $= 4.18  \times 10^{-5}$ kg/m.s, which is on the order of the value measured in laboratory experiments.
Equation \eqref{eq_sigma_5} predicts a pressure of 
$p = \rho U^2 = 3.1 \times 10^5$ Pa, which also is of the correct order of magnitude.  
Assuming $V = 1$ m/s and $L = 1$ m, we have $Re = V L / \nu = 6.9 \times 10^4$, $Kn = \lambda/L = 6.8 \times 10^{-8}$, $Re \, Kn = 4.7 \times 10^{-3} \ll 1$.  
Therefore, these numbers show that the above assumptions are reasonable and mutually acceptable.

\subsection{Evaluation of the Stress Tensor}
\label{sec_stress}
Our goal is to obtain closed-form expressions for the stress $\bs{\sigma}$ and force $\nabla\cdot\bs{\sigma}$ in terms of the hydrodynamic variables ($\rho$, $\ub$, $q$) for each of the Maxwell-Boltzmann, Cauchy, and general L\'evy $\alfa$-stable distributions.
Inserting \eqref{eq_f_general_soln_2} into \eqref{eq_sigma_def}, 
we have for the stress: 
\begin{align}
\sigma_{ij}{\ss(t,\x)} &= - \mediumiiint \mediumint (u_i - \bar{u}_i{\ss(t,\x)})(u_j - \bar{u}_j{\ss(t,\x)}) \,  f^{eq}_\alfa{\ss(t- s\tau,\x - s\tau\u,\u)} ~e^{-s} ~ ds \,d\u  \,.
\label{eq_sigma_def_4} 
\end{align}
The integrals in this expression would be straightforward to evaluate were it not for the temporal and spatial shifts in $f^{eq}_\alfa(t- s\tau,\x - s\tau\u,\u)$.  Indeed, with no shift, the time integral would decouple, and the stress would be:
\begin{equation}
 - \mediumiiint_{\!\!\!-\infty}^\infty (u_i - \bar{u}_i{\ss(t,\x)})(u_j - \bar{u}_j{\ss(t,\x)}) \,  f^{eq}_\alfa{\ss(t,\x,\u)} \,d\u   ~ \mediumint_0^\infty  ~e^{-s} ~ ds  = -p \, \delta_{ij}   
\label{eq_sigma_5}
\end{equation}
where $\delta_{ij}$ is the Kronecker delta, and $p = \rho U^2$ is the static pressure.   This result is identical to that obtained at the leading order of the {\it Chapman-Enskog expansion}, with the value $\rho U^2$ derived herein by carrying out the integrations in \eqref{eq_sigma_5}.  % 

To focus on the frictional shear stress, we decompose the total stress
as the sum of the {\it pressure} and  {\it deviatoric stress} in the usual way \citep{Kundu2012}
\be
\sigma_{ij}  = -p \, \delta_{ij} + \tau_{ij}   \,.
\label{eq_sigma_p_tau}
\ee
 Since $f^{eq}_\alfa{\ss(t- s\tau,\x - s\tau\u,\u)} = f^{eq}_\alfa{\ss(t,\x,\u)} + \{ f^{eq}_\alfa{\ss(t- s\tau,\x - s\tau\u,\u)} - f^{eq}_\alfa{\ss(t,\x,\u)} \}$, the
  deviatoric stress is 
\be
\begin{split}
\tau_{ij}{\ss(t,\x)} = - \mediumiiint \mediumint  (u_i - \bar{u}_i{\ss(t,\x)}) & (u_j - \bar{u}_j{\ss(t,\x)}) \, 
\big\{ f^{eq}_\alfa{\ss(t- s\tau,\x - s\tau\u,\u)} - f^{eq}_\alfa{\ss(t,\x,\u)} \big\} ~e^{-s} ~ ds \,d\u  
\label{eq_tau_1} 
\end{split}
\ee
or, using the notation from \eqref{eq_feq_F} for $f_\alfa^{eq}$:
\be
\begin{split}
\tau_{ij}{\ss(t,\x)} = - \frac{\rho}{U^3} \mediumiiint \mediumint & (u_i - \bar{u}_i{\ss(t,\x)})  (u_j - \bar{u}_j{\ss(t,\x)}) \, 
\\
& \quad\quad\quad \cdot
 \big\{ F(\Delta{\ss(t-s\tau,\x - s\tau\u,\u)}) - F(\Delta{\ss(t,\x,\u)}) \big\} ~e^{-s} ~ ds \,d\u  \,.
\label{eq_tau_2} 
\end{split}
\ee
The challenge in evaluating the stress \eqref{eq_tau_2} is in untangling $F(\Delta{\ss(t-s\tau,\x - s\tau\u,\u)})$ from appearing in both the $ds$ and $d\u$ integrals.  Two key insights are needed to untangle $F(\Delta{\ss(t-s\tau,\x - s\tau\u,\u)})$:  (1) since collisions occur on a much faster timescale than changes in the macroscopic flow, the time shift in $F(\Delta{\ss(t-s\tau,\,\ldots\,)})$ may be ignored; and (2) the $d\u$ integral in \eqref{eq_tau_2} can be made tractable by breaking it into two pieces, corresponding to ``small'' and ``large'' displacements.

\subsubsection{Removing the temporal shift}
\label{sec_remove_time_shift}
First, we demonstrate that the time-and-space-shifted $F(\Delta{\ss(t-s\tau,\x - s\tau\u,\u)})$ in \eqref{eq_tau_2} can be replaced by the space-only-shifted $F(\Delta{\ss(t,\x - s\tau\u,\u)})$.  
From the assumptions \eqref{eq_s_vs_unity} and \eqref{eq_time_vs_tau}, which prescribe $s \sim \O(1)$  and small $\tau$, we are authorized to perform a Taylor expansion in time:
\be
F(\Delta{\ss(t-s\tau,\x - s\tau\u,\u)})
= 
F(\Delta{\ss(t,\x - s\tau\u,\u)})
+
\frac{dF}{d\Delta} \frac{\p \Delta{\ss(t,\x - s\tau\u,\u)}}{\p t} (- s\tau) + \O(\tau^2) 
\label{eq_f_no_time}
\ee
where from \eqref{eq_feq_F}
\be
\frac{\p \Delta{\ss(t,\x - s\tau\u,\u)}}{\p t} = -\frac{2}{U^2} \text{\small $\sum\limits_{k=1}^3$}  (u_k - \bar{u}_k{\ss(t,\x- s\tau\u)})   \frac{\p \bar{u}_k{\ss(t,\x- s\tau\u)}}{\p t}   \,.
\label{eq_dDelta_dt}
\ee
Upon inserting \eqref{eq_f_no_time} and  \eqref{eq_dDelta_dt} into  \eqref{eq_tau_2}, it is evident that $dF/d\Delta$ always multiplies odd powers of velocity differences like $(u_i - \bar{u}_i)$.  Since $dF/d\Delta$ itself is even in such velocity differences, the $dF/d\Delta$ terms always integrate to zero, and \eqref{eq_tau_2} simplifies to
\be
\begin{split}
\tau_{ij}{\ss(t,\x)} = - \frac{\rho}{U^3} \mediumiiint \mediumint & (u_i - \bar{u}_i{\ss(t,\x)})  (u_j - \bar{u}_j{\ss(t,\x)}) \,
\big\{ F(\Delta{\ss(t,\x - s\tau\u,\u)}) - F(\Delta{\ss(t,\x,\u)}) \big\} ~e^{-s} ~ ds \,d\u  
\label{eq_tau_2t} 
\end{split}
\ee
with no longer a time shift in $F$, although the spatial shift remains.  

Since all terms are now evaluated at the present time $t$, we choose to drop from here onward the mention of $t$ among the arguments of all functions, with no implication of a steady state.

\subsubsection{Splitting of the spatial shift}
The key step in order to  untangle the $\int  ds$ and $\iiint  d\u$ integrals in \eqref{eq_tau_2t}
  is to split the $\iiint  d\u$ integral into two pieces, corresponding to ``small'' and ``large'' displacements ($\x-\x' = s\tau\u$).  Formally, we write
\be
\mediumiiint_{\!\!\!-\infty}^\infty  \, (\dots) \,d\u  
=
\underset{\!\!\! s\tau |\u| \leq \ell}{\mediumiiint_{\!\!-\infty}^\infty}   \, (\dots)\,d\u  
+
\underset{\!\!\!\!  s\tau |\u| \geq \ell}{\mediumiiint_{\!\!-\infty}^\infty}  \, (\dots)  \,d\u   
\label{eq_break} 
\ee
in which the demarcation distance $\ell$ is defined in \eqref{eq_ell} and is bracketed by \eqref{eq_L_scales}.

Since by virtue of \eqref{eq_L_scales}, the length $\ell$ is very short compared to the length scale $L$ of the flow, the first integral captures a minor redistribution of particles and may be identified as the contribution to the viscous shear stress, whereas the second integral corresponding to large displacements, up to the length scale of the flow, may be properly identified with a turbulent shear stress:
\be
\begin{split}
\tau_{ij}^{visc} \equiv
- \frac{\rho}{U^3} \underset{\!\!\! s\tau |\u| \leq \ell}{\mediumiiint} \mediumint & (u_i - \bar{u}_i{\ss(\x)})(u_j - \bar{u}_j{\ss(\x)}) \, 
 \big\{ F(\Delta{\ss(\x - s\tau\u,\u)}) - F(\Delta{\ss(\x,\u)}) \big\} ~e^{-s} ~ ds \,d\u   
 \label{eq_tau_visc}
\end{split}
\ee
\be
\begin{split}
\tau_{ij}^{turb}   \equiv
- \frac{\rho}{U^3} \underset{\!\!\!\!  s\tau |\u| \geq \ell}{\mediumiiint} \mediumint & (u_i - \bar{u}_i{\ss(\x)})(u_j - \bar{u}_j{\ss(\x)}) \, 
 \big\{ F(\Delta{\ss(\x - s\tau\u,\u)}) - F(\Delta{\ss(\x,\u)}) \big\} ~e^{-s} ~ ds \,d\u  ~.
\label{eq_tau_turb} 
\end{split}
\ee

For the viscous stress $\tau_{ij}^{visc}$, that for displacements $s\tau|\u| \le \mathcal{O}(\ell) \ll L$, we will 
perform Taylor expansions in space.  For the turbulent term, that for displacements 
$s\tau|\u| \ge \mathcal{O}(\ell)$, no such Taylor expansion is permitted, but we will be able 
to use an asymptotic expression of $\Delta$ and consider only the tails of the 
$f_\alpha^{eq}$ distribution.  Either way, the problem simplifies considerably.

\subsection{Small Displacements:  $s\tau |\u| \leq \ell$ }
\label{sec_small}

Consider first the case of small displacements $|\x-\x'| = s\tau|\u| \leq \ell$, wherein we seek to evaluate the viscous stress \eqref{eq_tau_visc}.
Since $\ell \ll L$ by virtue of \eqref{eq_L_scales}, the displacements are much smaller than the length scale $L$ over which the flow varies, and a Taylor expansion is permissible.  We write:
\be 
F(\Delta{\ss(\x - s\tau\u)}) = F(\Delta{\ss(\x)}) + \frac{dF(\Delta{\ss(\x)})}{d\Delta} \frac{\p\Delta{\ss(\x)}}{\p x_l}   (-s\tau u_l)   + \O((s\tau |\u|)^2)   
\label{eq_F_Taylor}
\ee
with implied sum over $l=1,2,3$.  Using \eqref{eq_Delta},  the spatial derivative of $\Delta$ is:
\be
\frac{\p\Delta}{\p x_l} = - \frac{2}{U^2} (u_k - \bu_k{\ss(\x)}) \frac{\p \bu_k}{\p x_l} 
\label{eq_dDelta_dx}
\ee
with implied sum over $k=1,2,3$.  
Now inserting \eqref{eq_F_Taylor} and \eqref{eq_dDelta_dx} into  \eqref{eq_tau_visc}, we obtain
\be
\tau_{ij}^{visc}
= - \frac{\rho}{U^3} \underset{\!\!\! s\tau |\u| \leq \ell}{\mediumiiint} \mediumint (u_i - \bar{u}_i)(u_j - \bar{u}_j) \,  
\left\{ \frac{dF}{d\Delta} \[ - \frac{2}{U^2} (u_k - \bu_k) \frac{\p \bu_k}{\p x_l} \]   (-s\tau u_l) \right\} ~e^{-s} ~ ds \,d\u  
\label{eq_I1_3} 
\ee
in which $i,j$ are fixed, summation is implied over $k,l = 1,2,3$, and all $\bu$ and $\Delta$ are evaluated at $\x$.  Note that since $dF/d\Delta$ is even with respect to any $(u_i - \bu_i)$,  only even terms survive the integral over $d\u$ (and only when  $u_l$ is shifted to $(u_l - \bu_l))$.  Thus, for $i\neq j$, the only terms that survive are $(k=i,l=j)$ and $(k=j,l=i)$.  
For $i=j$, all three $k=l$ terms survive.  

Since all terms are evaluated at $\x$, the only remaining coupling between the $s$ and $\u$ integrals is in the limits of the $\u$ integral.  Note that the $s$ integrand is exponentially small for large $s$, so Assumption \#2  (that $s \sim \O(1)$) is justified here.  Thus, it is reasonable to approximate the limit of the $\u$ integral as $|\u| \leq \ell/\tau$.  Then, the integral over $s$ has decoupled, and since $\int_0^\infty s \, e^{-s} ~ ds = 1$, we are left with
\be
\tau_{ij}^{visc} =
\begin{cases}
 -2 \frac{\rho \tau }{U^5} \[  \frac{\p \bu_i}{\p x_j} +  \frac{\p \bu_j}{\p x_i} \] \underset{\!\!\! |\u| \leq \ell/\tau}{\iiint}  (u_i - \bar{u}_i)^2(u_j - \bar{u}_j)^2 \,  
 \frac{dF}{d\Delta}   \,d\u  
&(i \neq j) \\[1 em]
-2 \frac{\rho \tau }{U^5}   \frac{\p \bu_i}{\p x_i}  \underset{\!\!\! |\u| \leq \ell/\tau}{\iiint}  (u_i - \bar{u}_i)^4 \,  
 \frac{dF}{d\Delta}   \,d\u    
\\ \quad\quad\quad
-2 \frac{\rho \tau }{U^5}  \sum\limits_{k = 1 \atop k\neq i}^3  \frac{\p \bu_k}{\p x_k} \underset{\!\!\! |\u| \leq \ell/\tau}{\iiint}  (u_i - \bar{u}_i)^2(u_k - \bar{u}_k)^2 \,  
\frac{dF}{d\Delta}  \,d\u  
&(i = j)
\end{cases}
\label{eq_I1_4} 
\ee
with no summation implied over $i,j$. 

We can make some additional headway in simplifying \eqref{eq_I1_4} by realizing that $dF/d\Delta$ is only a function of the velocity magnitude, so the integrals in  \eqref{eq_I1_4} can be converted into spherical coordinates, and the angle integrals can be evaluated without specification of $dF/d\Delta$.  Using spherical coordinates ($r,\theta,\phi)$ such that $(u-\bu,v-\bv,w-\bw) = (r\,\sin\phi\,\cos\theta,r\,\sin\phi\,\sin\theta,r\,\cos\phi)$, $d(u-\bu)\,d(v-\bv)\,d(w-\bw) = r^2 \, \sin\phi \, d\phi\,d\theta\,dr$, and $\Delta = r^2/U^2$, consider the following two prototypical integrals:
\begin{equation} \begin{split}
I_{22} &\equiv
\underset{\!\!\! |\u| \leq \ell/\tau}{\mediumiiint}
(u - \bar{u})^2(v - \bar{v})^2 \,  
 \frac{dF}{d\Delta}    \,d(\u - \ub) 
= 
\frac{4\pi}{15}
\int_0^{\ell/\tau} r^6 
 \frac{dF}{d\Delta}    \,dr  \\
I_4 &\equiv
\underset{\!\!\! |\u| \leq \ell/\tau}{\mediumiiint}
(u - \bar{u})^4 \,  
\frac{dF}{d\Delta}    \,d(\u - \ub)    
= 
\frac{4\pi}{5}
\int_0^{\ell/\tau} r^6 
 \frac{dF}{d\Delta}    \,dr 
=
3 I_{22}  ~.
\end{split} \end{equation}
Since $I_4 = 3 I_{22}$, assuming incompressibility ($\sum_{k=1}^3 \frac{\p \bu_k}{\p x_k}  = 0$), the $i=j$ case in  \eqref{eq_I1_4} simplifies to be identical in form to the $i\neq j$ case.  Therefore,  \eqref{eq_I1_4} can be written as follows for any $i,j$
\be
\tau_{ij}^{visc}{\ss(\x)} =
 -2\frac{\rho \tau }{U^5}  \[  \frac{\p \bu_i}{\p x_j} +  \frac{\p \bu_j}{\p x_i} \] I_{22} 
 \quad\quad , \quad\quad
 I_{22} = 
\frac{4\pi}{15}
\int_0^{\ell/\tau} r^6 
 \frac{dF}{d\Delta}    \,dr ~.
\label{eq_I1_5} 
\ee
Evaluation of the remaining integral in \eqref{eq_I1_5} depends on the choice made for the equilibrium distribution $F$.  

\subsubsection{Maxwell-Boltzmann distribution ($\alfa=2$)}
With the {\it Maxwell-Boltzmann distribution} \eqref{eq_F_MB}, 
$F(\Delta) = (2\pi)^{-3/2} e^{-\Delta/2}$, 
$dF/d\Delta = -\half (2\pi)^{-3/2} e^{-\Delta/2}$, with $\Delta = r^2/U^2$.  Since this integrand of $I_{22}$ decays exponentially and $\ell/\tau$ is very large, it is reasonable to approximate $I_{22}$ by extending the integral to infinity.  Then, $I_{22} \approx -\half U^7$, and the stress is 
\be
\tau_{ij}^{visc}{\ss(\x)}  = \mu \[  \frac{\p \bu_i}{\p x_j}  + \frac{\p \bu_j}{\p x_i} \],  \quad\quad    \mu = \rho U^2\tau
\label{eq_tau_visc_final}
\ee
which is the stress for a Newtonian fluid.  This end result is identical to that obtained via the {\it Chapman-Enskog expansion}, which also assumes small $\tau$ and a Maxwell-Boltzmann distribution \citep{Chapman1991}.

Note that the transport variable in the Maxwell-Boltzmann ($\alfa = 2$) case is the kinematic viscosity, $\nu \equiv \mu/\rho$, which is here shown to be $\nu = U^2 \tau$.

\subsubsection{Cauchy distribution ($\alfa=1$)}
If instead we select the {\it Cauchy distribution} \eqref{eq_F_C},  
$F(\Delta)  =  \pi^{-2}  (\Delta + 1)^{-2}$, then 
$dF/d\Delta =  -2\pi^{-2} (\Delta + 1)^{-3}$, and
\begin{equation} \begin{split}
I_{22} &= \frac{4\pi}{15} \int_0^{\ell/\tau} r^6  \frac{-2}{\pi^2 (\Delta + 1)^3}  \,dr  
=   \frac{-8 U^6}{15 \pi}  \int_0^{\ell/\tau}   \frac{r^6 \, dr}{(r^2 + U^2)^3}  \\
&=
\frac{-8 U^6}{15 \pi}
\[ \frac{8 r^5 + 25 r^3 U^2 + 15 r U^4}{8 (r^2 + U^2)^2} - \frac{15 U}{8}\tan^{-1}\frac{r}{U}   \]_{r=0}^{\tfrac{\ell}{\tau}} 
\approx  \frac{-8 U^6}{15 \pi} \frac{\ell}{\tau}
\label{eq_I1_5C2} 
\end{split} \end{equation}
where the last simplification was made possible because $\tfrac{\ell}{\tau} \gg U$.  Inserting \eqref{eq_I1_5C2} into \eqref{eq_I1_5}, the  stress then is 
\be
\tau_{ij}^{visc}{\ss(\x)}  =   \frac{16}{15 \pi} \rho U \ell   \[  \frac{\p \bu_i}{\p x_j}  + \frac{\p \bu_j}{\p x_i} \]  ~.
\label{eq_I1_5C3a}
\ee

Note that although \eqref{eq_I1_5C3a} appears to be an eddy viscosity model, in fact it should still be interpreted as the stress due to molecular viscosity.  The viscosity in  \eqref{eq_I1_5C3a}, 
$\mu_1 = \frac{16}{15 \pi} \rho  U \ell$, 
is different than that in \eqref{eq_tau_visc_final}, $\mu$, because the Cauchy distribution of molecular velocities ascribes different agitation to the fluid.  
It is expected that the form of \eqref{eq_tau_visc_final} will hold true for any ($0 < \alfa < 2$), with viscosity $\mu_\alfa$ depending on the choice of $\alfa$ as in \eqref{eq_I1_5C3a}. 

\subsubsection{Friction force (small displacements)}
The friction force  appearing in the momentum equation is then obtained by taking the divergence of the stress \eqref{eq_tau_visc_final} or \eqref{eq_I1_5C3} in the usual way:
\be
(\nabla\cdot\bs{\TAU}^{visc})_j{\ss(\x)} = \frac{\p \tau_{ij}^{visc}{\ss(\x)}}{\p x_i}  
= \mu_\alfa \[  \frac{\p^2 \bu_i}{\p x_i \p x_j}  + \frac{\p^2 \bu_j}{\p x_i^2} \]  ~.
\ee
Assuming incompressibility $\frac{\p \bu_i}{\p x_i} = 0$, the first term is zero, leaving 
\be
(\nabla\cdot\bs{\TAU}^{visc})_j{\ss(\x)} 
= \mu_\alfa  \nabla^2 \bu_j  ~,
\label{eq_forcing_small_displacements}
\ee
which is the usual result for a Newtonian fluid.

\subsection{Large Displacements: $s\tau |\u| \geq \ell$}
\label{sec_large}
We now consider the case of large displacements, $s\tau |\u| \geq \ell$, wherein we seek to evaluate the turbulent stress \eqref{eq_tau_turb}, which is repeated here for convenience:
\be
\tau_{ij}^{turb}{\ss(\x)}
\equiv 
- \frac{\rho}{U^3} \underset{\!\!\! s\tau |\u| \geq \ell}{\mediumiiint} \mediumint (u_i - \bar{u}_i{\ss(\x)})(u_j - \bar{u}_j{\ss(\x)}) \,  \big\{ F(\Delta{\ss(\x - s\tau\u)}) - F(\Delta{\ss(\x)}) \big\} ~e^{-s} ~ ds \,d\u  ~.
\label{eq_tau_turb} 
\ee
To begin, recall that large displacements $s\tau |\u| \geq \ell$ imply large speeds $|\u| \gg |\ub|$, which permits the neglect of 
the $\bar{u}_i$ and $\bar{u}_j$ terms: 
\be
\tau_{ij}^{turb}{\ss(\x)}
=
- \frac{\rho}{U^3} \underset{\!\!\! s\tau |\u| \geq \ell}{\mediumiiint} \mediumint u_i u_j  \,  \big\{ F(\Delta{\ss(\x - s\tau\u)}) - F(\Delta{\ss(\x)}) \big\} ~e^{-s} ~ ds \,d\u  ~.
\label{eq_I2_2} 
\ee
Our strategy now is to decouple the time and velocity integrals with the following substitution,
$u_i = (x_i-x_i')/(s\tau)$ and $d\u = d\x'/(-s\tau)^3$, which replaces velocity with its corresponding displacement over flight time:
\be
\tau_{ij}^{turb}{\ss(\x)}
=
- \frac{\rho}{U^3} \underset{\!\!\!\! |\x'-\x|  \geq \ell}{\mediumiiint} \mediumint \frac{x_i-x_i'}{s\tau} \frac{x_j-x_j'}{s\tau}  \,  \big\{ F(\Delta{\ss(\x')}) - F(\Delta{\ss(\x)}) \big\} ~e^{-s} ~ ds \,\frac{d\x'}{(s\tau)^3}   ~.
\label{eq_I2_3} 
\ee
The negative signs in the $d\x'/(-s\tau)^3$ terms have been used to flip the integration limits in \eqref{eq_I2_3}.

Note that time still appears implicitly in the $F(\Delta)$ terms (by virtue of the velocities therein), so the time and space integrals are not yet decoupled.  To proceed, recall that large displacements $s\tau |\u| \geq \ell$ also imply $\Delta \gg 1$, so we need only consider the tails of $F(\Delta)$ when evaluating \eqref{eq_I2_3}.  Moreover, since $|\u| \gg |\ub|$, we can further simplify the form of $\Delta$ and $F(\Delta)$ as follows:
Rewriting definition \eqref{eq_Delta}, we have
\begin{align}
\Delta{\ss(\x)} &= \text{\scriptsize $\sum\limits_{k=1}^3$} \Big\{ \tfrac{u_k^2}{U^2} - 2 \tfrac{u_k \bu_k{\ss(\x)}}{U^2} +  \tfrac{\bar{u}_k{\ss(\x)}^2}{U^2} \Big\}  \equiv A - 2 B + C
\label{eq_feq_F_2}  \\
\Delta{\ss(\x')} &= \text{\scriptsize $\sum\limits_{k=1}^3$} \Big\{ \tfrac{u_k^2}{U^2} - 2 \tfrac{u_k \bu_k{\ss(\x')}}{U^2} +  \tfrac{\bar{u}_k{\ss(\x')}^2}{U^2} \Big\} \equiv A - 2 B' + C'  ~.
\label{eq_feq_F_2p}  
\end{align}
Since $|\u| \gg |\ub|$, we note the ordering $A \gg B \gg C$ and $A \gg B' \gg C'$.  Further, since the difference  $\Delta{\ss(\x')} - \Delta{\ss(\x)} \approx 2 (B-B')$ is much smaller than $\Delta$ itself (since $\Delta \approx A$), the difference $F(\Delta{\ss(\x')}) - F(\Delta{\ss(\x)})$ may be simplified via Taylor series expansion:
\be
\begin{split}
F(\Delta{\ss(\x')}) - F(\Delta{\ss(\x)}) 
&\approx \frac{dF(\Delta{\ss(\x)})}{d\Delta}  (\Delta{\ss(\x')} - \Delta{\ss(\x)})
\approx \frac{dF(\Delta{\ss(\x)})}{d\Delta}  2 (B-B') \\
&= 2 \frac{dF(\Delta{\ss(\x)})}{d\Delta}   \frac{\u \cdot (\ub{\ss(\x)} - \ub{\ss(\x')})}{U^2} ~.
\label{eq_dFdDelta}
\end{split}
\ee
Upon inserting \eqref{eq_dFdDelta} into \eqref{eq_I2_3} and making the substitution $\u = (\x-\x')/(s\tau)$, we have
\be
\begin{split}
\tau_{ij}^{turb}{\ss(\x)}
=
 \frac{\rho}{U^5 \tau^6} \underset{\!\!\!\! |\x'-\x|  \geq \ell}{\iiint}& \int  (x_i'-x_i) (x_j'-x_j)  \, 
\bigg\{ 2 \tfrac{dF}{d\Delta}   (\x'-\x) \cdot (\ub{\ss(\x)} - \ub{\ss(\x')})  \bigg\} ~e^{-s} \,\frac{ds}{ s^6} d\x'   ~.
\end{split}
\label{eq_I2_4b}
\ee
Further progress can only be made by specifying the equilibrium distribution $F(\Delta)$.

Note how at this stage the stress expression \eqref{eq_I2_4b}, originally quadratic in velocity differences, has morphed into an expression that is {\it linear} in velocity differences.  The quadratic factors have now been expressed in terms of displacements, while a velocity difference re-emerged from the Taylor expansion of the distribution function.

\subsubsection{Maxwell-Boltzmann distribution ($\alfa=2$)}
The tails of the Maxwell-Boltzmann distribution fall of exponentially, so the turbulent stress $\tau_{ij}^{turb}$ is effectively zero.  To show this formally, recall that for the Maxwell-Boltzmann distribution, 
$F(\Delta) = \frac{1}{(2\pi)^{3/2}} e^{-\Delta/2}$, so 
\be
\frac{dF}{d\Delta} = -\frac{1}{2} \frac{1}{(2\pi)^{3/2}} e^{-\Delta/2} \approx -\frac{1}{2} \frac{1}{(2\pi)^{3/2}} e^{-|\u|^2/2 U^2} = -\frac{1}{2} \frac{1}{(2\pi)^{3/2}} e^{-|\x-\x'|^2/2 (s\tau U)^2} ~.
\ee
Thus, for the Maxwell-Boltzmann distribution, the dimensionless time part of the integral 
in \eqref{eq_I2_4b} is 
\begin{equation}
I = \int_0^\infty ~e^{-\half D s^{-2} - s} ~\frac{ds}{s^6} ~.
\label{eq_tau_large_MB}
\end{equation}
where we have defined $D\equiv |\x-\x'|^2/(U\tau)^2$ to simplify the algebra.
This integral can be approximated in order to have an idea of its size.  A change of variable 
$\beta = 1/s^5$ transforms \eqref{eq_tau_large_MB} into
\begin{equation}
I = \frac{1}{5} \int_0^\infty e^{-\half D \beta^{2/5} - \beta^{-1/5}} ~d\beta ~.
\label{eq_tau_large_MB_2}
\end{equation}
This integrand decays to zero rapidly for $\beta$ away from both sides of the $\beta_0$ that minimizes the negative of the exponent,
\begin{equation}
\half D \beta^{2/5} + \beta^{-1/5}  ~.
\label{eq_exponent}
\end{equation}
This (negative) exponent reaches a minimum (of $\frac{3}{2}D^{1/3}$) at  $\beta_0 = D^{-5/3} = (|\x-\x'|/(U\tau))^{-10/3}$.  An estimate of the integral can be made by expanding the exponent  \eqref{eq_exponent} in a  
Taylor series (with $\beta = \beta_0 + \bar{\beta}$ for small $\bar{\beta}$):
\begin{equation}
\half D \beta^{2/5} + \beta^{-1/5} 
= \tfrac{1}{2} \beta_0^{-3/5} (\beta_0 + \bar{\beta})^{2/5} + (\beta_0 + \bar{\beta})^{-1/5}
= \tfrac{3}{2} \beta_0^{-1/5} + \tfrac{3}{50} \beta_0^{-11/5} \bar{\beta}^2 + O(\bar{\beta}^3) .
\label{eq_beta_taylor}
\end{equation}
Retaining only the leading terms and inserting the expanded exponent \eqref{eq_beta_taylor} into \eqref{eq_tau_large_MB_2}, we obtain
\begin{equation}
I = \frac{1}{5} ~e^{-\frac{3}{2} \beta_0^{-1/5}} 
\int_{-\beta_0}^\infty e^{-\frac{3}{50} \beta_0^{-11/5} \bar{\beta}^2} ~d\bar{\beta} 
\le 
\sqrt{\frac{2\pi}{3}} ~\beta_0^{11/10} ~e^{-\frac{3}{2} \beta_0^{-1/5}} ~,
\end{equation}
with the upper bound obtained by extending the lower limit of integration all the way 
to $-\infty$.  The factor $\beta_0^{11/10} = (|\x-\x'|/U\tau)^{-11/3}$ is small while 
the exponent $-\frac{3}{2}\beta_0^{-1/5} = -\frac{3}{2}(|\x-\x'|/U\tau)^{2/3}$ is very large negative 
making the exponential vanishingly small.

The conclusion is that the Maxwell-Boltzmann distribution with its rapid exponential tail 
does not contribute in any significant way to the stress tensor in the range of large 
displacements.  In other words, the Maxwell-Boltzmann distribution is associated 
with laminar viscous stress only and produces no turbulent stress.  The fact that the 
Maxwell-Boltzmann distribution in Boltzmann kinetics leads to the Navier-Stokes equations 
has been known for a long time \citep{Chapman1991}.

One element must be clear.  The absence of  large-displacement, turbulent-like frictional 
stress with the use of the Maxwell-Boltzmann distribution is not equivalent to stating 
that the choice of a Maxwell-Boltzmann distribution precludes the presence of turbulence.  
It only says that all turbulent fluctuations then arise from other (advective) terms in 
the momentum equations.  Turbulence is kept explicit and unresolvable (by direct numerical simulation for high-Reynolds number flows) even with today's 
computer power.  What we will show next is that the consideration of large displacements 
with a heavy-tail distribution allows for the capture of (at least some) turbulent 
fluctuations in some statistical way without need to resolve them in all their details.

\subsubsection{General L\'evy $\alfa$-stable distribution ($\alfa < 2$)}
\label{sec_tau_turb_Levy}

The Cauchy distribution is but one member of the family of {\it L\'evy $\alpha$-stable distributions}, that for $\alpha = 1$.  Since the tail behavior of the Cauchy distribution (and hence the derivation of the turbulent stress) follows that of the general case, we proceed by deriving the turbulent stress for the general case of any $0 < \alpha < 2$.  

Recall, our starting point is equation \eqref{eq_I2_4b}, which is repeated here for convenience
\be
\begin{split}
\tau_{ij}^{turb}{\ss(\x)}
=
 \frac{\rho}{U^5 \tau^6} \underset{\!\!\!\! |\x'-\x|  \geq \ell}{\iiint}& \int  (x_i'-x_i) (x_j'-x_j)  \, 
\bigg\{ 2 \tfrac{dF}{d\Delta}   (\x'-\x) \cdot (\ub{\ss(\x)} - \ub{\ss(\x')})  \bigg\} ~e^{-s} \,\frac{ds}{ s^6} d\x'   ~.
\end{split}
\tag{\ref{eq_I2_4b}}
\ee
Our immediate task is to evaluate the derivative $dF/d\Delta$ within the tails of $F(\Delta)$.  
The tails of the 3D multivariate L\'evy $\alfa$-stable distributions behave 
asymptotically as \citep{Nolan2006}
\begin{align}
F(\Delta) &\simeq  \frac{\bar{C}_\alfa}{\Delta^{(\alfa+3)/2}}   \quad\quad \text{for} \quad \Delta \gg 1,  \,\,
   \text{ where} \quad
\bar{C}_\alfa \equiv 
\frac{2^\alfa \, \Gamma\(\frac{\alfa + 3}{2}\)}{\pi^\frac{3}{2}  \big|\Gamma\(-\frac{\alfa}{2}\)\big|}  ~.
\label{eq_Levy_tails}
\end{align}
The derivative of $F$ is asymptotically 
\begin{equation}
\frac{dF}{d\Delta} \simeq -\frac{\alpha + 3}{2} ~\frac{\bar{C}_\alpha}{\Delta^{(\alpha+5)/2}} \quad\quad \text{for} \quad \Delta \gg 1 ~.
\label{eq_dFdDelta_levy}
\end{equation}
For example, the Cauchy distribution ($\alfa=1$) has tails $F(\Delta) \simeq \frac{1}{\pi^2}\frac{1}{\Delta^2}$ and asymptotic derivative $dF/d\Delta \simeq -\frac{2}{\pi^2}\frac{1}{\Delta^3}$.

Inserting \eqref{eq_dFdDelta_levy} into \eqref{eq_I2_4b}, and with the asymptotic expression $\Delta \approx A = |\u|^2/U^2 = |\x - \x'|^2/(s\tau U )^2$, we have
\be
\begin{split}
\tau_{ij}^{turb} 
= 
 \frac{\rho}{U^5 \tau^6}
\underset{\!\!\!\! |\x'-\x|  \geq \ell}{\iiint} \int 
 (x_i'-x_i) & (x_j'-x_j)  \,  
\bigg\{ (\alpha + 3)  \bar{C}_\alpha ~\frac{(s\tau U)^{\alpha+5}}{|\x'-\x|^{\alpha+5}}  
 \\[-0.5 em]
&\quad\quad\quad
\cdot
(\x'-\x) \cdot (\ub{\ss(\x')} - \ub{\ss(\x)})  \bigg\} ~e^{-s} ~ \frac{ds}{s^6} \,d\x' ~.
\end{split}
\label{eq_tau_3L3} 
\ee 
or upon simplification
\be
\begin{split}
\tau_{ij}^{turb} 
= 
 \rho\tfrac{ (U\tau)^\alfa}{\tau} (\alpha + 3)  \bar{C}_\alpha &
\underset{\!\!\!\! |\x'-\x|  \geq \ell}{\iiint} \!\! \int
 (x_i'-x_i) (x_j'-x_j)  \,  
\frac{(\x'-\x) \cdot (\ub{\ss(\x')} - \ub{\ss(\x)})}{ |\x'-\x|^{\alpha+5} } \frac{e^{-s} ds}{s^{1-\alfa}} \, d\x' .
\end{split}
\label{eq_tau_3L4} 
\ee 
The time integral now decouples, and its value is  $\int_0^\infty s^{\alfa - 1} e^{-s} ds = \Gamma(\alfa)$.  
Thus, for ``large'' displacements we have finally
\be
\boxed{
\tau_{ij}^{turb}{\ss(\x)} =  \rho\tfrac{ (U\tau)^\alfa}{\tau} (\alpha + 3) \Gamma(\alfa) \bar{C}_\alfa     \underset{\!\!\!\! |\x'-\x|  \geq \ell}{\iiint}    (x_i'-x_i) (x_j'-x_j)  \,      \frac{ (\x'-\x)  \cdot (\ub{\ss(\x')} - \ub{\ss(\x)})}{|\x'-\x|^{\alfa+5} }   \,d\x'  
}~.
\label{eq_tau_5L} 
\ee
Appendix \ref{sec_alternate_tau_turb} provides an alternate form of the turbulent stress \eqref{eq_tau_5L}, which some may find useful.

\subsubsection{Friction force (large displacements)}
The friction force appearing in the momentum equation is the divergence of this stress tensor, $(\nabla\cdot\bs{\tau}^{turb})_j{\ss(\x)} = \tfrac{\p \tau_{ij}^{turb}{\ss(\x)}}{\p x_i}$. 
Taking derivatives of \eqref{eq_tau_5L} term by term, and summing over $i=1,2,3$, we have:
\begin{equation}
\begin{split}
(\nabla\cdot\bs{\tau}^{turb})_j{\ss(\x)}  
=  \rho \tfrac{ (U\tau)^\alfa}{\tau}  
& (\alpha + 3)  \Gamma(\alfa) \bar{C}_\alfa  \\
\cdot \underset{\!\!\!\! |\x'-\x|  \geq \ell}{\iiint}     \text{\scriptsize $\sum\limits_{i=1}^3$} \Big\{ 
&-  (x_j'-x_j)  \, \frac{ (\x'-\x) \cdot (\ub{\ss(\x')} - \ub{\ss(\x)})}{|\x'-\x|^{\alpha + 5}} \\
&-  (x_i'-x_i) \delta_{ij}  \, \frac{ (\x'-\x) \cdot (\ub{\ss(\x')} - \ub{\ss(\x)})}{|\x'-\x|^{\alpha + 5}}  \\
&-  (x_i'-x_i)  (x_j'-x_j)  \, \frac{ \bu_i{\ss(\x')} - \bu_i{\ss(\x)}}{|\x'-\x|^{\alpha + 5}} \\
&+ (\alpha + 5)  (x_i'-x_i)^2 (x_j'-x_j)  \, \frac{ (\x'-\x) \cdot (\ub{\ss(\x')} - \ub{\ss(\x)})}{|\x'-\x|^{\alfa+7}}  \\
& -  (x_i'-x_i) (x_j'-x_j) (x_k'-x_k) \, \frac{\p \bu_k{\ss(\x)}}{\p x_i} \, \frac{ 1}{|\x'-\x|^{\alpha + 5}} 
\Big\} \,d\x'  
\label{eq_forcing_1L} 
\end{split}
\end{equation}
with summation implied over $k=1,2,3$.  
The first four terms combine (with $-3-1-1 + (\alfa+5) = \alfa$), and the fifth term integrates to zero (since it is odd), leaving
\begin{equation}
(\nabla\cdot\bs{\tau}^{turb})_j{\ss(\x)}  
= \rho \tfrac{ (U\tau)^\alfa}{\tau} 
\alfa (\alpha + 3) \Gamma(\alfa) \bar{C}_\alfa \underset{\!\!\!\! |\x'-\x|  \geq \ell}{\iiint}    
  (x_j'-x_j)  \, \frac{ (\x'-\x) \cdot (\ub{\ss(\x')} - \ub{\ss(\x)})}{|\x'-\x|^{\alpha + 5}} 
\,d\x'  ~.
\label{eq_forcing_2L} 
\end{equation}
Equation \eqref{eq_forcing_2L}  can be written as a fractional Laplacian upon integrating by parts: 
$ \int a \, db = ab - \int b \, da $.  
First, write the dot product out into three terms 
$(x_k'-x_k) (\bu_k{\ss(\x')} - \bu_k{\ss(\x)})$, with implied summation over $k=1,2,3$.  Now, for each of the three terms, perform partial integration in $dx'_k$ 
\begin{align*}
a  &=  (x_j'-x_j) (\bu_k{\ss(\x')} - \bu_k{\ss(\x)})  ~,
&
db &= \frac{ x_k'-x_k }{|\x'-\x|^{\alpha + 5}}  \,dx_k' ~, \\
da &= \Big\{  \delta_{jk} (\bar{u}_k{\ss(\x')} - \bar{u}_k{\ss(\x)})    + (x_j'-x_j) \tfrac{\p \bu_k{\ss(\x')}}{\p x_k'} \Big\} dx_k'  ~,
&
b &=  \frac{  -1 }{(\alpha + 3) |\x'-\x|^{\alpha + 3}} ~,
\end{align*}
with no summation implied in each partial integration (fixed $k$).  
Since $ab\big|_{-\infty}^\infty = 0$ for each of these partial integrations (so long as $|\ub{\ss(\x')}| / |\x'|^{\alfa+2} \ra 0 $ as $|\x'| \ra \infty$), the other terms can be recombined (by now implying summation over $k$):
\be
\begin{split}
(\nabla\cdot\bs{\tau}^{turb})_j{\ss(\x)}  
= \rho \tfrac{ (U\tau)^\alfa}{\tau}  
\Gamma(\alfa{+}1) \bar{C}_\alfa  \underset{\!\!\!\! |\x'-\x|  \geq \ell}{\iiint}    
  \frac{  \delta_{jk} (\bar{u}_k{\ss(\x')} - \bar{u}_k{\ss(\x)})    + (x_j'-x_j) \tfrac{\p \bu_k{\ss(\x')}}{\p x_k'} }{|\x'-\x|^{\alpha + 3}}  d\x'  .
\label{eq_forcing_3L} 
\end{split}
\ee
where we used the identity $\alfa \Gamma(\alfa) = \Gamma(\alfa{+}1)$.
For incompressible flow, $\tfrac{\p \bu_k}{\p x_k'} = 0$, so this term is eliminated.  
Note that $\ell/L = Re^{-3/4} \ra 0$ as $Re \ra \infty$, so \eqref{eq_forcing_3L} is asymptotically equivalent to a {\it Cauchy principal value integral}.  Thus, we write
\begin{equation}
\boxed{
(\nabla\cdot\bs{\tau}^{turb})_j{\ss(\x)}  
=  \rho \frac{(U\tau)^\alfa}{\tau} 
\Gamma(\alfa{+}1) \bar{C}_\alfa  
\dashint\!\!\!\dashint\!\!\!\dashint_{\!-\infty}^\infty
  \frac{  \bar{u}_j{\ss(\x')} - \bar{u}_j{\ss(\x)}  }{|\x'-\x|^{\alfa+3}} 
\,d\x'  
} ~.
\label{eq_forcing_fractional_Laplacian} 
\end{equation}
Within a coefficient of proportionality, Equation \eqref{eq_forcing_fractional_Laplacian} prescribes the turbulent force as the ``singular integral'' form of the {\it fractional Laplacian}. 
Indeed, the fractional Laplacian of function $g(\x)$ is defined as \citep{Kwasnicki2017}
\begin{equation}
(-\nabla^2)^{\frac{\alpha}{2}} g(\x) 
= \bar{C}_\alfa
\dashint\!\!\!\dashint\!\!\!\dashint_{\!-\infty}^\infty
       \frac{g(\x') - g(\x)}{|\x'-\x|^{\alpha + 3}} ~d\x'
\label{eq_fractional_Laplacian_dD}
\end{equation}
with the same constant $\bar{C}_\alfa$ as defined in \eqref{eq_Levy_tails}. 
Note that the derivation herein was presented for 3D space, and the derivation for 1D or 2D space follows similarly, with the constant $\bar{C}_\alfa$ and exponents adjusted accordingly.  Appendix \ref{sec_order_reduction} shows the reduction in order of the fractional Laplacian  if the flow is two dimensional.

\subsection{Momentum Equation}
\label{sec_momentum_eqn}
The momentum equation can now be formulated by inserting the sum of the small-displacement forcing \eqref{eq_forcing_small_displacements} and large-displacement forcing \eqref{eq_forcing_fractional_Laplacian} into momentum equation \eqref{eq_momentum}
\begin{equation}
\boxed{
\rho \frac{D\ub}{Dt}
= -\nabla p + \mu_\alfa \nabla^2 \ub + \rho C_\alfa  
\dashint\!\!\!\dashint\!\!\!\dashint_{\!-\infty}^\infty
    \frac{  \ub{\ss(\x')} - \ub{\ss(\x)}  }{|\x'-\x|^{\alfa+3}} 
\,d\x'
}
\label{eq_momentum_evaluated}
\end{equation}
where 
$C_\alfa \equiv  \tfrac{(U\tau)^\alfa}{\tau} \Gamma(\alfa{+}1) \bar{C}_\alfa$ 
is a {\it turbulent mixing coefficient}, which has units (length)$^\alfa$/(time).  This is the main result, as given in Equation \eqref{eq_momentum_0}.  

Note that the Navier-Stokes equations are recovered from \eqref{eq_momentum_evaluated} if the equilibrium distribution is assumed to be Maxwell-Boltzmann.  With $\alfa=2$, we have $\bar{C}_\alfa = 0$ and $\mu_\alfa = \mu$, so the turbulent contribution is zero, and the viscous force is that of a Newtonian fluid.

\subsection{Properties of the Fractional Laplacian}
\label{sec_properties}
The friction force appearing in the momentum Equation \eqref{eq_momentum_evaluated} is 
\be
{\bf F}{\ss(\x)} =  \rho C_\alfa   
\dashint\!\!\!\dashint\!\!\!\dashint_{\!-\infty}^\infty
    \frac{  \ub{\ss(\x')} - \ub{\ss(\x)}  }{|\x'-\x|^{\alfa+3}}   \,d\x'  ~.
\label{eq_force_general}
\ee
In this section, we consider some properties of this operator.

\subsubsection{Conservation of momentum}
The fractional Laplacian respects conservation of momentum, since its spatial integral across the entire domain vanishes
\be
\iiint\limits_{\mathbb{R}^3}  {\bf F}{\ss(\x)} \, d\x =
\iiint\limits_{\mathbb{R}^3} \rho C_\alfa      
\dashint\!\!\!\dashint\!\!\!\dashint_{\!-\infty}^\infty
    \frac{  \ub{\ss(\x')} - \ub{\ss(\x)}  }{|\x'-\x|^{\alfa+3}}  \,d\x'  \,d\x 
    = 0  ~.
\ee
The proof is obvious from the symmetry of the integrand; substitute $\x'$ for $\x$ and vice versa, and find that the integral is equal to minus itself.  Thus, a net momentum gain or loss occurs only at the boundaries.  

\subsubsection{Energy dissipation}
The fractional Laplacian  dissipates kinetic energy. The rate of work done on the fluid (per unit volume) is ${\bf F}{\ss(\x)} \cdot \ub{\ss(\x)}$, so the total work rate for the whole fluid domain is
\begin{align}
\iiint\limits_{\mathbb{R}^3}  {\bf F}{\ss(\x)} \cdot \ub{\ss(\x)} \, d\x &=
\iiint\limits_{\mathbb{R}^3} \rho C_\alfa     
\dashint\!\!\!\dashint\!\!\!\dashint_{\!-\infty}^\infty
    \frac{  \ub{\ss(\x')} - \ub{\ss(\x)}  }{|\x'-\x|^{\alfa+3}} \cdot \ub{\ss(\x)} \,d\x'  \,d\x  \nonumber \\
&=
-\iiint\limits_{\mathbb{R}^3} \rho C_\alfa      
\dashint\!\!\!\dashint\!\!\!\dashint_{\!-\infty}^\infty
    \frac{  \ub{\ss(\x')} - \ub{\ss(\x)}  }{|\x'-\x|^{\alfa+3}} \cdot \ub{\ss(\x')} \,d\x'  \,d\x    \\
&=
-\half \iiint\limits_{\mathbb{R}^3} \rho C_\alfa      
\dashint\!\!\!\dashint\!\!\!\dashint_{\!-\infty}^\infty
    \frac{  |\ub{\ss(\x')} - \ub{\ss(\x)}|^2  }{|\x'-\x|^{\alfa+3}}  \,d\x'  \,d\x  \leq 0  ~,  \nonumber
\end{align}
where the first equality comes from the anti-symmetry of the integrand, and the second equality comes from averaging the two previous expressions. 
Since the work rate is negative (energy is being dissipated), the total kinetic energy can only decrease over time (unless energy is provided at the boundaries).

\section{Comparison with Turbulent Transport Theory}

At this stage, we developed two parallel formalisms, one for the diffusion of a passive scalar (\S\ref{sec_transport}) and the other for the friction force acting on momentum 
(\S\ref{sec_derivation}).  Each equation we obtained has a fractional Laplacian term representing the effect of  turbulent motions: %  
\begin{align}
\frac{D\bar{c}}{Dt}&=   q \gamma^\alpha  \bar{C}_\alpha
\dashint\!\!\!\dashint\!\!\!\dashint_{\!-\infty}^\infty
   \frac{ \bar{c}{\ss(t,\x')} -  \bar{c}{\ss(t,\x)} }{|\x' - \x|^{\alpha+3}}  
      d\x'
\tag{\ref{eq_transport_general_final}} \\
 \frac{D\ub}{Dt}
&= -\tfrac{1}{\rho} \nabla p + \tfrac{\mu_\alfa}{\rho} \nabla^2 \ub 
+ \tfrac{(U\tau)^\alfa}{\tau}  
\Gamma(\alfa{+}1) \bar{C}_\alfa    
\dashint\!\!\!\dashint\!\!\!\dashint_{\!-\infty}^\infty
    \frac{  \ub{\ss(t,\x')} - \ub{\ss(t,\x)}  }{|\x'-\x|^{\alfa+3}} 
\,d\x' ~.
\tag{\ref{eq_momentum_evaluated}}
\end{align}
Clearly, the two expressions share a similar structure with a fractional Laplacian, and the parallelism invites the equating of the front coefficients:
\be
\frac{(U\tau)^\alfa}{\tau} \Gamma(\alfa{+}1)  = \gamma^\alpha  q ~,
\label{eq_Utau_to_q}
\ee
where parameter $\gamma$ may change values with $\alfa$. 

This is an important relation, because it links the microscale variables $U$ and $\tau$ of Boltzmann kinetics to the macroscale quantities $q = \nu$, $u_*$, or $\epsilon$ of the observable world.  For the special values of $\alfa$ (Table~\ref{tab_similarity}), we obtain:
\begin{align}
\alfa &= 2: 			&	2 U^2\tau &= \gamma^2 \nu   
\label{eq_U_vs_q_alfa2}  \\
\alfa &= 1: 			&	U	    &= \gamma u_*   
\label{eq_U_vs_q_alfa1}\\
\alfa &= \tfrac{2}{3}:  &  \Gamma\(\tfrac{5}{3}\)  U^\frac{2}{3} \tau^{-\frac{1}{3}}  &=  \gamma^\frac{2}{3} \epsilon^\frac{1}{3} ~.  \label{eq_U_vs_q_alfa23}
\end{align}

The undetermined factor $\gamma$ is not surprising, considering that a passive scalar is typically not dispersed at exactly the same rate as momentum; the ratio of the scalar diffusivity to momentum diffusivity (kinematic viscosity) is the Prandtl number, which is typically not unity.  So, we surmise that there ought to be a connection between $\gamma$ and the Prandtl number.

\subsection{Remark on the Molecular Viscosity in Shear Turbulence ($\alfa=1$)}
With $U$ set to $\gamma\, u_*$ using \eqref{eq_U_vs_q_alfa1}, we may recast  \eqref{eq_I1_5C3a} only in terms of macroscopic quantities:
\be
\tau_{ij}^{visc}{\ss(\x)}  =   \mu_1  \[  \frac{\p \bu_i}{\p x_j}  + \frac{\p \bu_j}{\p x_i} \], \quad\quad 
\mu_1 \equiv \frac{16}{15 \pi} \rho \gamma\, u_* \ell   ~.
\label{eq_I1_5C3}
\ee 
Taking $\ell$ to be the Kolmogorov microscale, then 
$\mu_1 / \mu = \frac{16}{15 \pi} \rho \gamma \, u_* (L Re^{-3/4}) / \mu
= \frac{16}{15 \pi} \gamma \frac{u_*}{V}  Re^{1/4}$.  For flow past a flat plate at $Re = 10^6$, this ratio is $\mu_1/\mu = 0.51 \gamma$, 
which is $\O(1)$ as expected.

\subsection{Form of the Momentum Equation for $\alfa = 1$ and 2/3}
For shear turbulence, Table \ref{tab_similarity} suggests assuming a Cauchy distribution, with $\alfa=1$, $U = \gamma\,u_*$, and $\bar{C}_1 = 1/\pi^2$.  In this case,  
and with \eqref{eq_I1_5C3}, Equation \eqref{eq_momentum_evaluated} becomes
\begin{equation}
\rho \frac{D\ub}{Dt}
= -\nabla p +  \frac{16}{15 \pi} \rho  \,\gamma\, u_*  \ell\, \nabla^2 \ub + \frac{ \rho  \, \gamma\, u_*}{\pi^2}   
\dashint\!\!\!\dashint\!\!\!\dashint_{\!-\infty}^\infty
    \frac{  \ub{\ss(\x')} - \ub{\ss(\x)}  }{|\x'-\x|^4} 
\,d\x'  ~.
\label{eq_momentum_Cauchy}
\end{equation}

For intertial turbulence, Table \ref{tab_similarity} suggests assuming $\alfa = 2/3$ and $q = \eps^{1/3}$. 
The coefficient of the fractional Laplacian \eqref{eq_Levy_tails} evaluates to $\bar{C}_{\frac{2}{3}} = 0.0660$.
The value of the enhanced viscosity $\mu_\alfa$ and parameter $\gamma$ for the $\alfa = 2/3$ case are not known at this time, but the momentum equation has the following struture
\be
\rho \frac{D\ub}{Dt}
= -\nabla p + \mu_\alfa \nabla^2 \ub + 0.0660 \rho \gamma^{2/3} \eps^{1/3} 
\dashint\!\!\!\dashint\!\!\!\dashint_{\!-\infty}^\infty
    \frac{  \ub{\ss(\x')} - \ub{\ss(\x)}  }{|\x'-\x|^{11/3}} 
\,d\x'  ~.
\ee

%----------------------------------------------------------------------------------------------------
%----------------------------------------------------------------------------------------------------
%----------------------------------------------------------------------------------------------------
%----------------------------------------------------------------------------------------------------
%----------------------------------------------------------------------------------------------------
%----------------------------------------------------------------------------------------------------
%----------------------------------------------------------------------------------------------------
%----------------------------------------------------------------------------------------------------
%----------------------------------------------------------------------------------------------------
%----------------------------------------------------------------------------------------------------
%----------------------------------------------------------------------------------------------------
%----------------------------------------------------------------------------------------------------
\section{Examples}
\label{sec_examples}
We preface this section by noting that the derivation above only strictly applies to unbounded flows.   
Rigorous treatment of boundary conditions is rather involved and is beyond the scope of this article.  In lieu of that protracted derivation, herein we simply truncate the fractional Laplacian to an integration over the fluid domain.  Thus, the results in this section are intended to be a preliminary assessment of the fractional Laplacian, with proper treatment of the boundary conditions appearing in a future publication.

\subsection{Logarithmic velocity profile in the Cauchy case}
\label{sec_log_profile}

Consider 1D flow in a semi-infinite domain ($z \geq 0$), with velocity field $\ub = [\bu{\ss(z)},0,0]$.  Assume $\alfa=1$ consistent with the Cauchy distribution, and consider the limit of extremely high Reynolds number.  With these assumptions, the fractional Laplacian can immediately be reduced to 1D (see Appendix \ref{sec_order_reduction}),
\be
F_x{(z)} = \frac{ \rho \,\gamma \, u_*}{\pi}  {\dashint_0^\infty}  ~ \frac{  \bu{\ss(z')} - \bu{\ss(z)}  }{(z-z')^2}  \,dz'  ~,
\label{eq_LLWC_1}
\ee
and the momentum equation \eqref{eq_momentum_evaluated} reduces to $F_x{(z)} = 0$.
Ignoring any boundary conditions (\ie allowing the fluid to slip along the wall), the exact solution of $F_x{(z)} = 0$ is the logarithmic profile, $\bu{\ss(z)} = \ln(z)$, consistent with the {\it Law of the Wall} \eqref{eq_Law_of_the_Wall}.
For proof,  split the integral into two portions,
\be
{\dashint_0^\infty} ~ \frac{ \ln(z') - \ln(z)  }{(z-z')^2}  \,dz' = \int_0^{z^-} (\cdots) dz' + \int_{z^+}^\infty (\cdots) dz'  \nonumber
\ee
and then change variables $z' = z^2/\zeta$ in the second integral, $dz' = - z^2/\zeta^2 \, d\zeta$.  The second integral is then found to be equal and opposite to the first:
\be
\begin{split}
\frac{\pi}{ \rho \,\gamma\,u_*} F_x(z)  &=   \int_0^{z^-} \frac{\ln(z') - \ln(z)}{(z'-z)^2} dz'
-  \int_{z^-}^0 \frac{ \ln(z^2/\zeta) - \ln(z)}{(z^2/\zeta-z)^2} \frac{ z^2 }{\zeta^2} \, d\zeta  \\ 
%&
&=   \int_0^{z^-} \frac{\ln(z'/z)}{(z'-z)^2} dz'
-  \int_0^{z^-} \frac{\ln(\zeta/ z) }{(z - \zeta)^2} \, d\zeta = 0 ~.
\label{eq_Fx_is_zero_for_log_profile}
\end{split}
\ee
Thus,
the logarithmic profile would be the leading-order solution of a turbulent flat plate boundary layer problem away from the boundaries.  This strongly suggests that the case $\alfa = 1$ with the Cauchy distribution corresponds to wall turbulence.

\subsection{Couette Flow}
Couette flow is shear-driven flow between parallel plates extending to infinity in the $x$ and $y$ directions.  Thus, the ensemble-averaged flowfield is $\ub = (\bu{\ss(z)},0,0)$, with boundary conditions $u{\ss(z=0)} = 0$ and $u{\ss(z=L)} = V$, where $L$ is the gap height, and $V$ is the speed of the top plate. The pressure is uniform. 
The Reynolds number is defined as $Re = \rho VL/\mu$ and is assumed to be large.  We further assume a turbulent flow that is steady in the mean.

Assuming the random speeds $\u$ follow a Cauchy $f^{eq}_\alfa$ distribution, the momentum equation reduces to
\begin{equation}
0
=  \frac{16}{15 \pi} \rho \,\gamma \, u_* \ell \, \frac{d^2 \bu}{d z^2} + \frac{ \rho \,\gamma \, u_*}{\pi^2}   
\dashint_0^L \dashint_{-\infty}^\infty \dashint_{-\infty}^\infty   ~   
    \frac{  \bu{\ss(z')} - \bu{\ss(z)}  }{[(x'-x)^2 + (y'-y)^2 + (z'-z)^2]^2} 
\,dx'\,dy' \,dz'   ~.
\label{eq_momentum_Couette}
\end{equation}
Since the velocity $\bu{\ss(z)}$ is not a function of $x$ or $y$, the fractional Laplacian can be reduced from 3D to 1D by carrying out the $x$ and $y$ integrations (see Appendix \ref{sec_order_reduction}):
\begin{equation}
0
=  \frac{16}{15 \pi} \rho \,\gamma \, u_* \ell \, \frac{d^2 \bu}{d z^2} + \frac{ \rho \,\gamma \, u_*}{\pi} 
\dashint_0^L
    \frac{  \bu{\ss(z')} - \bu{\ss(z)}  }{(z'-z)^2} 
 \,dz'
\label{eq_Couette_2}
\end{equation}
Finally, we nondimensionalize $\bu^* = \bu/V$ and $z^* = z/L$ but immediately drop the stars to obtain
\begin{equation}
0
=  \frac{16}{15} \frac{\ell}{L} \frac{d^2 \bu}{d z^2} +   
\dashint_0^1
    \frac{  \bu{\ss(z')} - \bu{\ss(z)}  }{(z'-z)^2} 
 \,dz'
\label{eq_Couette_4}
\end{equation}
with non-dimensional boundary conditions $u{\ss(z=0)} = 0$ and $u{\ss(z=1)} = 1$.

Note that for infinite Reynolds number ($\ell/L = Re^{-3/4} \ra 0$) and ignoring the boundary conditions (\ie allowing wall slip), the solution of \eqref{eq_Couette_4} is the double-log profile, $\bu(z) = \ln[z/(1-z)]$.  The proof is similar to that in \S\ref{sec_log_profile}:
\be
F_x(z)  =  \dashint_0^1 \frac{ \ln[z'/(1-z')] - \ln[z/(1-z)]}{(z'-z)^2} dz' % 
= \int_0^{z^-} \frac{\ln\frac{z' (1-z)}{z(1-z')} }{(z'-z)^2} dz' 
     +\int_{z^+}^1 \frac{\ln\frac{z' (1-z)}{z(1-z')} }{(z'-z)^2} dz' ~. \nonumber 
\ee
Upon making the substitution $z'/(1-z') = [z/(1-z)]^2/[\zeta/(1-\zeta)]$, $dz' = - \{ z'(1-z')/[\zeta(1-\zeta)] \} \, d\zeta$, the second integral is found (after some tedious algebra) to be equal and opposite to the first
\be
\begin{split}
F_x(z)  &= \int_0^{z^-} \frac{  \ln\frac{z' (1-z)   }{z(1-z')   }  }{(z'-z)^2} dz' 
         - \int_{z^-}^0 \frac{  \ln\frac{z (1-\zeta)}{\zeta(1-z)}  }{(z'-z)^2}  \frac{ z'(1-z')}{\zeta(1-\zeta)} \, d\zeta  \\
&=         \int_0^{z^-} \frac{  \ln\frac{z' (1-z)   }{z(1-z')   }  }{(z'-z)^2} dz'
        -  \int_0^{z^-} \frac{  \ln\frac{\zeta(1-z)}{z (1-\zeta)}  }{(z - \zeta)^2} \, d\zeta = 0  ~.
\label{eq_Fx_is_zero_for_double_log_profile}
\end{split}
\ee
Thus, for the case of turbulent Couette flow a double-log profile is analytically predicted for the core region.

Now at a finite Reynolds number ($\ell/L = Re^{-3/4}$), 
Equation \eqref{eq_Couette_4} can be discretized and solved numerically.  Upon defining 
$\Delta z = 1/(N-1)$, 
$z_i = (i-1) \Delta z$ for $i=1,\ldots,N$, 
$D \equiv \frac{16}{15} Re^{-3/4}/\Delta z^2$, 
and weights $W_{ij} = (1-\delta_{ij}) \Delta z/(z_j-z_i)^2$, 
we can write \eqref{eq_Couette_4} as 
\be
D (\bu_{i+1} - 2 \bu_i + \bu_{i-1}) + \sum_{j=1}^N W_{ij}  (\bu_j - \bu_i) = 0 
\label{eq_Couette_4numerical}
\ee
with boundary conditions $\bu_1 = 0$ and $\bu_N = 1$.

Since $\ell/L \ll 1$, the molecular friction force is of lower order than the turbulent friction force, except perhaps in a laminar sublayer of height proportional to $\ell/L$.  Thus, our expectation is that solution of \eqref{eq_Couette_4numerical} will yield a log profile that is desingularized at the wall.  One such desingularized double-log profile is (in non-dimensional terms) 
\be
\bu(z) = \frac{1}{2} - \frac{1}{2} \frac{\ln[(d+z)/(d+1-z)]}{\ln[d/(d+1)]}  ~,
\label{eq_u_desingular_double_log_profile}
\ee
where $d$ is a small number ($d\ll 1$) that represents a viscous sublayer or roughness height.
Equation \eqref{eq_u_desingular_double_log_profile} simultaneously meets the boundary conditions and has the proper asymptotic behavior away from the walls $d \ll z \ll 1-d$.

\begin{figure}
\begin{center}
\includegraphics[width=3.25 in]{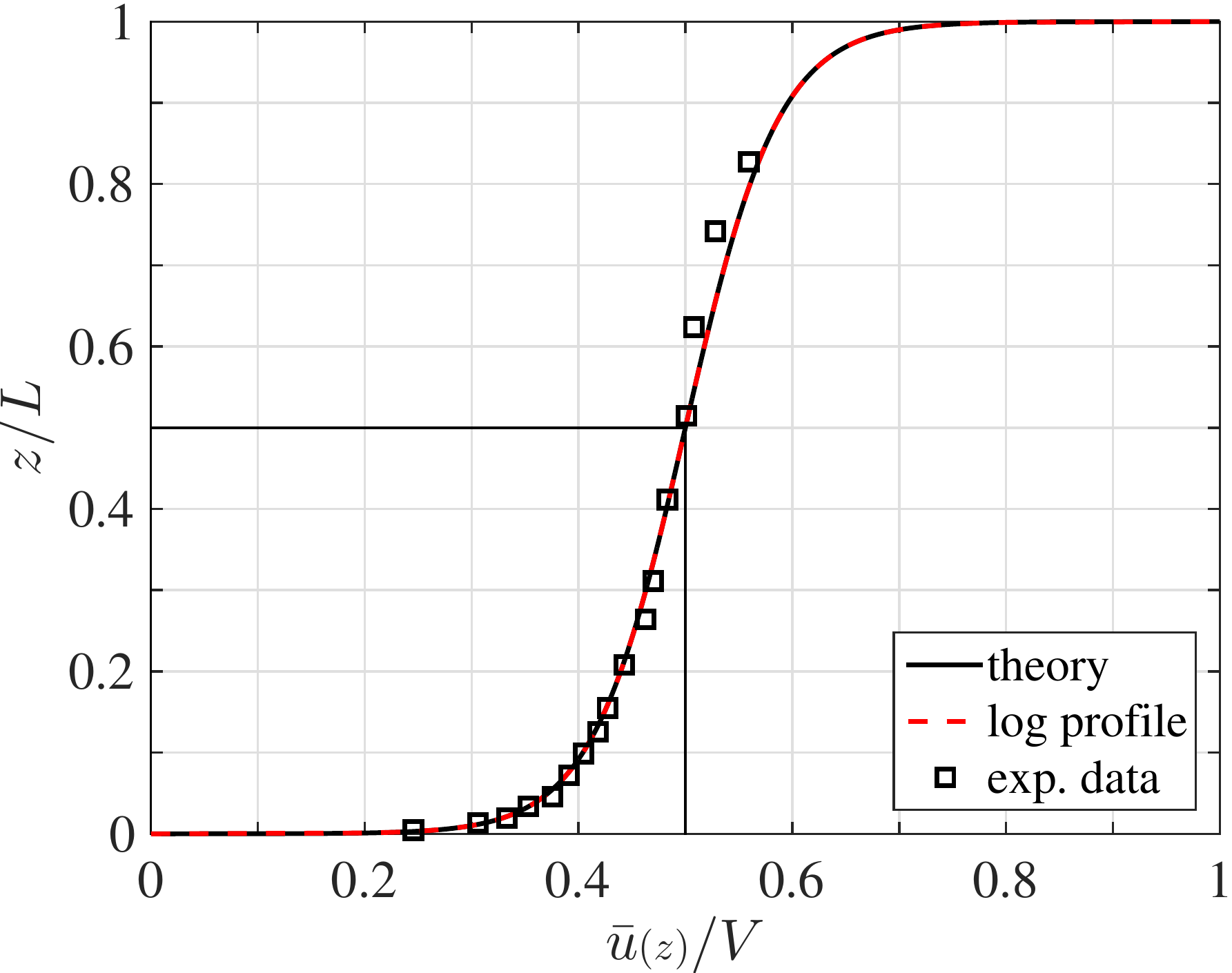}
%\vspace{-7 mm}
\caption{Couette flow velocity profile: `--' numerical solution of \eqref{eq_Couette_4numerical} with $N = 3200$ grid nodes; ``-\,-'' best fit log profile with $d = 1.06 \times 10^{-5}$ \eqref{eq_u_desingular_double_log_profile}; `$\Box$' experimental data for flow at $Re = 16,500$ \citep{Robertson1970}.
}
\label{fig_Couette_results}
\end{center}
\end{figure}

Figure \ref{fig_Couette_results} shows the velocity profile predicted by numerical solution of \eqref{eq_Couette_4numerical}, as well as the best fit log profile \eqref{eq_u_desingular_double_log_profile} and experimental data from \citep{Robertson1970}.  The log profile fits the numerical solution very well, and the agreement with the experiment is very encouraging.  We emphasize, however, that these results are tentative, because of our tentative treatment of the boundary conditions for the fractional Laplacian.  Rigorous treatment of the boundary conditions remains a formidable task and will be addressed in a subsequent publication.  

\subsection{2D Boundary Layer Flow}
Consider now a 2D turbulent boundary layer flow over a flat plate, with plate length $L$ 
and freestream speed $V$.  Assume the boundary layer thickness, $\delta{\ss(x)}$, is small $\delta{\ss(x)}/L \ll 1$ and consequently $\frac{\p }{\p x} \ll \frac{\p }{\p z}$.  Assume $\alfa = 1$ consistent with a Cauchy equilibrium distribution.  After integration over transverse coordinate $y$ in the fractional Laplacian, the momentum equation \eqref{eq_momentum_Cauchy} reduces to
\begin{equation}
\bu \frac{\p\bu}{\p x} + \bw \frac{\p\bu}{\p z} 
=  \frac{16}{15 \pi} \,\gamma \,  u_* \ell \, \frac{\p^2\bu}{\p z^2} 
+ 
\frac{ \gamma \, u_*}{2 \pi} 
 \dashint_0^\infty   \dashint_{-\infty}^\infty 
    \frac{  \bu{\ss(x',z')} - \bu{\ss(x,z)}  }{[(x'-x)^2 + (z'-z)^2]^{3/2}}  \,dx' \,dz'  ~.
\label{eq_momentum_BL}
\end{equation}
The approximation $\frac{\p }{\p x} \ll \frac{\p }{\p z}$ permits a Taylor expansion for $\bu{\ss(x',z')}$
 such that
\begin{equation}
\bu \frac{\p\bu}{\p x} + \bw \frac{\p\bu}{\p z} 
=  
\frac{16}{15 \pi}  \,\gamma \, u_* \ell \, \frac{\p^2\bu}{\p z^2} 
+
\frac{ \gamma \,  u_* }{2 \pi} 
 \dashint_0^\infty   \dashint_{-\infty}^\infty     \frac{  \bu{\ss(x,z')} - \bu{\ss(x,z)}  + \frac{\p \bu{\ss(x,z')}}{\p x} (x' - x)   }{[(x'-x)^2 + (z'-z)^2]^{3/2}}  \,dx' \,dz'  ~.
\label{eq_momentum_BL2}
\end{equation}
The derivative term integrates to zero by symmetry.   The remaining terms in the numerator have no dependence on $x'$, so the $x'$ integral can be carried out trivially. For a very high Reynolds number, the viscous term is negligible, and the result is (along with the incompressible continuity equation)
\begin{equation}\begin{split}
  \frac{\p\bu}{\p x} +  \frac{\p\bw}{\p z}  &= 0 \\
 \bu \frac{\p\bu}{\p x} + \bw \frac{\p\bu}{\p z} 
&=  
\frac{ \gamma \, u_* }{\pi} 
\dashint_0^\infty
    \frac{  \bu{\ss(x,z')} - \bu{\ss(x,z)}  }{(z'-z)^2}   \,dz'  ~.
\label{eq_momentum_BL3}
\end{split}
\end{equation}

Equation \eqref{eq_momentum_BL3} is now parabolic in $x$ and can be solved numerically by the usual ``space marching'' method, where the solution at each discrete $x_i$ propagates forward to the next station, $x_{i+1}$.  Here, we discretize 
$x_i = \frac{i-1}{N_x-1} L$ and $z_j = \frac{j-1}{N_z-1} H$ for 
$i=1,\ldots,N_x$, $j=1,\ldots,N_z$, 
$N_x = 200$, and $N_z = 3000$; we choose $Re = 10^6$, $V = 1$ m/s, $L = 1$ m, and simulation domain height $H = 0.11$ m (which is about $9 \delta{\ss(x=L)}$). 
The advection terms (left hand side) of equation \eqref{eq_momentum_BL3} are discretized using a finite volume formulation with a linear upwind scheme \citep{Versteeg2007}.  The right hand side of \eqref{eq_momentum_BL3} was discretized as in \eqref{eq_Couette_4numerical} with 
weight matrix 
$W_{ij} = \frac{\gamma \, u_*}{\pi} (1-\delta_{ij}) \Delta z/(z_j-z_i)^2$ and solved numerically. 

Experimental observations of turbulent boundary layer flow all collapse to the
{\it Law of the Wall}:
\be
\frac{\bu{\ss(x,z)}}{u_*{\ss(x)}} = \frac{1}{\kappa} \ln\(\frac{z \, u_*{\ss(x)} }{ \nu}\) +  C^+ 
\label{eq_Law_of_the_Wall}
\ee
where $u_* \equiv \sqrt{\tau_w/\rho}$ is the {\it friction velocity}, $\tau_w{\ss(x)}$ is the wall shear stress, $\kappa = 0.41$ is the {\it von K\'arm\'an constant}, 
and $C^+ = 5.0$ is a constant \citep{Schlichting2000}.  
The friction velocity $u_*{\ss(x)}$ was computed herein from \citep{Schlichting2000} Equations [2.13], [17.60], and [18.99].
By definition, 
$Re_x \equiv V x / \nu = Re \cdot x/L$,   
$\Lambda{\ss(x)} \equiv \ln(Re_x)$, and  % Schlichting (18.99)
$D \equiv 2 \ln \kappa + \kappa(C^+ - 3.0)$.  % Schlichting (18.99)
Then, function $G{\ss(x)}$ is determined from solution of implicit equation
$\frac{\Lambda}{G} + 2\ln \frac{\Lambda}{G} - D = \Lambda$.   % Schlichting (17.60)
These data yield the {\it friction velocity}  
$u_*{\ss(x)} = V   \kappa \, G{\ss(x)}/\Lambda{\ss(x)}$  % [m/s] friction velocity,  Schlichting (18.76) 
and velocity profile $\bu{\ss(x,z)}$ \eqref{eq_Law_of_the_Wall}.  
The 99\% boundary layer thickness $\delta{\ss(x)}$ was computed numerically by interpolation of 
$\bu{\ss(x,z)}$ 
and is nearly equal to
$\delta{\ss(x)} \approx 0.11\, x \, G{\ss(x)}/\Lambda{\ss(x)} = 0.11\, x \, u_*{\ss(x)} / (V \kappa)$.   
At the plate end, $\delta{\ss(x=L)} =  0.0129$ m and $u_*{\ss(x=L)} = 0.0479$ m/s.

With $u_*$ set to a constant value of 0.0479 m/s in \eqref{eq_momentum_BL3}, 
the best match between the numerical solution of \eqref{eq_momentum_BL3} and the law of the wall \eqref{eq_Law_of_the_Wall} was found by setting $\gamma = 0.192$.  Physically, the parameter $\gamma$ controls the amount of momentum transport into the plate and the boundary layer growth rate in the numerical solution of \eqref{eq_momentum_BL3}. While we find $\gamma=0.192$ produces a good fit here, this value remains to be tested in other applications.

\begin{figure}
\begin{center}
\includegraphics[width= 0.95 \textwidth]{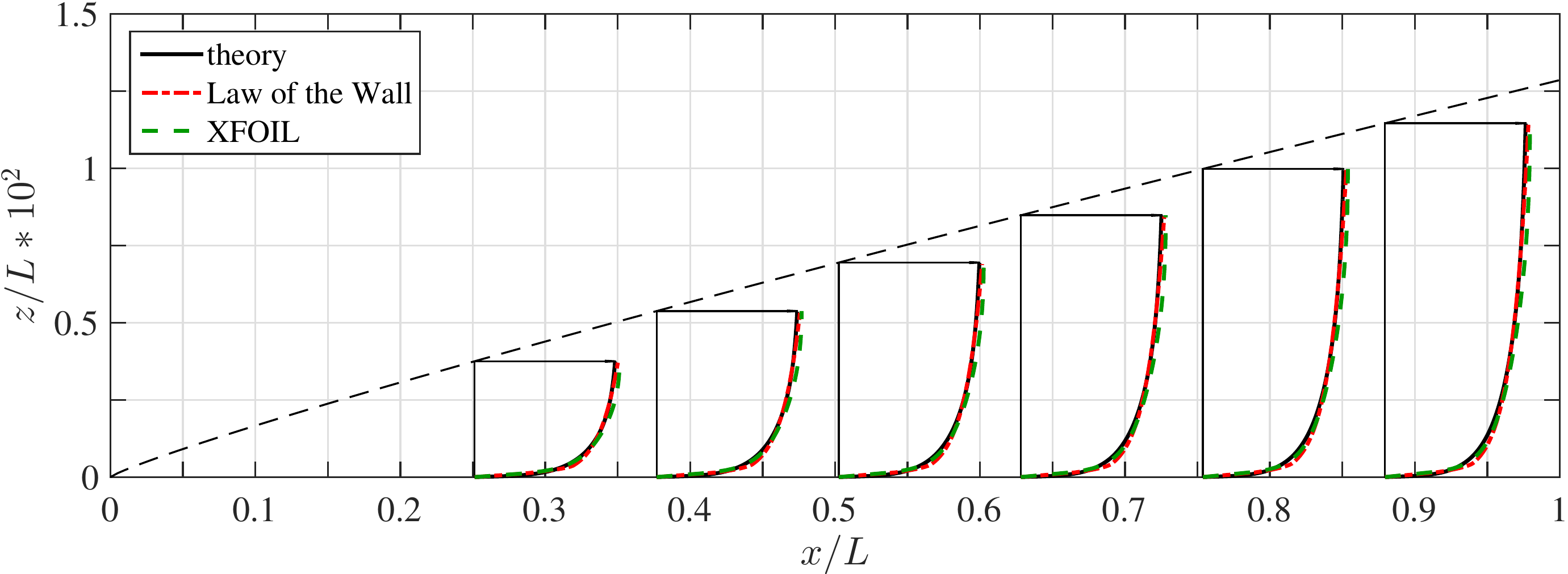}
\caption{2D boundary layer velocity profile: `--' numerical solution of \eqref{eq_momentum_BL3} with $N_x = 200$ and $N_z = 3000$ grid points; `-.-' {\it Law of the Wall} \eqref{eq_Law_of_the_Wall} with $\kappa = 0.41$ and $C^+ = 5.0$  \citep{Schlichting2000}; `-\,-' velocity profiles predicted by \texttt{XFOIL} subroutine \texttt{UWALL}, which blends the Law of the Wall with the Spalding formula for the outer layer \citep{Drela1989}.
}
\label{fig_BL}
\end{center}
\end{figure}

Encouragingly, the numerical solution of Equation \eqref{eq_momentum_BL3} well resembles the logarithmic velocity profile predicted by the Law of the Wall \eqref{eq_Law_of_the_Wall}, as shown in Figure \ref{fig_BL}. Figure \ref{fig_BL} also shows
the velocity profile predicted by using the Spalding formula for the outer layer, which is done in the widely-used airfoil analysis program \texttt{XFOIL} \citep{Drela1989}.  As with the Couette results, we remind the reader that these results are preliminary, because our treatment of the boundary conditions for the fractional Laplacian is tentative.  

We justify the use of the Cauchy $f^{eq}_\alfa$ distribution in this problem by the fact that the boundary layer thickness grows nearly linearly with respect to distance along the plate $\delta \sim x$.  This scaling can be seen in Figure~\ref{fig_BL}, where the boundary layer thickness is shown as a dashed line.  Since $G{\ss(x)}/\Lambda{\ss(x)}$ asymptotes to nearly constant value, $u_*{\ss(x)}$ is nearly constant, and $\delta{\ss(x)} \sim  x \, u_*/V = u_* t$, where $t = x/V$ is the freestream flight time.  Note that $\delta$ represents a distance of momentum diffusion, so an appropriate similarity variable describing turbulent transport is $\eta = \delta / (u_* t)$, which has linear scaling with time, consistent with $\alfa = 1$.  So, once again, we conclude that the choice $\alfa = 1$ is particularly well suited for the modeling of wall turbulence.

% Results from:  171115_2DZPreg_Re1e6_U2Cf_Nz3000.mat

%----------------------------------------------------------------------------------------------------
%----------------------------------------------------------------------------------------------------
%----------------------------------------------------------------------------------------------------
%----------------------------------------------------------------------------------------------------
%----------------------------------------------------------------------------------------------------
%----------------------------------------------------------------------------------------------------
%----------------------------------------------------------------------------------------------------
%----------------------------------------------------------------------------------------------------%----------------------------------------------------------------------------------------------------
%----------------------------------------------------------------------------------------------------
%----------------------------------------------------------------------------------------------------
%----------------------------------------------------------------------------------------------------
\section{Conclusions}
\label{sec_conclusions}

Starting from the framework of Boltzmann kinetic theory and calculating its associated stress field, we showed that the friction force in the momentum equations may be conveniently split into a first contribution 
from small displacements, yielding a molecular viscous component, and a second contribution 
from large displacements, yielding a fractional Laplacian component.  The demarcation 
length that separates small from large displacements was chosen as the Kolmogorov microscale.  Thus, the fractional Laplacian can be viewed as the part of the friction force that is generated by 
turbulent motions.

Boltzmann kinetic theory presumes the existence of an equilibrium distribution for the velocity 
fluctuations, which is traditionally taken as the Maxwell-Boltzmann (Gaussian) distribution.  
Through consideration of turbulent transport, we showed here that this traditional choice is not necessary and that any L\'evy $\alfa$-stable 
distribution is an acceptable alternative.  This family of distributions is characterized 
by a dimensionless parameter $\alpha$, which ranges from $0 < \alfa \leq 2$.  Among the possible 
values of $\alpha$, three stand out.  
For $\alpha = 2$, the equilibrium distribution is the traditional Maxwell-Boltzmann, and 
friction is accomplished solely by molecular viscosity.  
For $\alpha = 1$, the equilibrium distribution is the Cauchy distribution, and the fractional Laplacian reproduces the logarithmic velocity profile of a turbulent flow along
a wall. The formalism is also shown to properly represent the growth of a turbulent boundary layer along a plate and its accompanying logarithmic velocity profile (Law of the
Wall). 
In the case $\alpha = \frac{2}{3}$, the formalism should have the potential to represent inertial 
turbulence (Kolmogorov cascade), but this has not yet been verified.

Thus, the selection of the $\alpha$ value must reflect the nature of the flowfield 
under consideration: $\alpha = 2$ for laminar flows controlled by molecular viscosity $\nu$, 
$\alpha = 1$ for shear turbulence characterized by a friction velocity $u_*$, and 
$\alpha = 2/3$ for inertial turbulence characterized by an energy dissipation $\epsilon$.  
A formalism suitable for the calculation of a flowfield with all three processes acting 
simultaneously remains to be developed.

The fractional Laplacian formalism raises a few ancillary questions, the answer to which 
requires additional work.  First, a physical connection should exist between the ``thermal" 
agitation energy $\frac{3}{2}\rho U^2$ of Boltzmann kinetics and the turbulent kinetic energy.  
A further line of inquiry is whether Boltzmann kinetic energy could 
also provide an equation governing the evolution of the turbulent kinetic energy.

The fractional Laplacian is formally defined as an integration over a spatial domain that 
is infinite in all three dimensions.  But, as spatial dimensions are necessarily finite 
in practice, boundary conditions need to be imposed.  Proper treatment of these boundary conditions 
remains an open question, so as a tentative approach, herein we simply truncated the spatial 
integration.  The results for wall turbulence are most encouraging.

%----------------------------------------------------------------------------------------------------
%----------------------------------------------------------------------------------------------------
%----------------------------------------------------------------------------------------------------
%----------------------------------------------------------------------------------------------------
%----------------------------------------------------------------------------------------------------
\section*{Acknowledgements}
The authors are grateful to Dr.\ Hudong Chen of Exa Corporation for conversations and encouragement that helped shape this work.

%----------------------------------------------------------------------------------------------------
%----------------------------------------------------------------------------------------------------
%----------------------------------------------------------------------------------------------------%----------------------------------------------------------------------------------------------------
%----------------------------------------------------------------------------------------------------
%----------------------------------------------------------------------------------------------------
%----------------------------------------------------------------------------------------------------
\appendix

\section{Demarcation Scale $\ell$ as the Kolmogorov Microscale}
\label{sec_Kolmogorov}

Kolmogorov's theory of the turbulence cascade (``intertial turbulence'') is centered on 
the existence of a scale-independent transfer of energy, from the largest scale where 
turbulence is stirred by one or several instability mechanisms down to the shortest scale 
where viscous dissipation takes hold.  The {\it energy dissipation} (rate of kinetic energy dissipation per mass) is denoted $\epsilon$ and has the 
dimensions of energy per mass per time.

From elementary dimensional analysis, it follows that given any intermediate length scale 
(``eddy diameter'') $d$, the associated turbulent velocity fluctuations $u$ are given by
\begin{equation}
u = A ~(\epsilon d)^{1/3} \simeq (\epsilon d)^{1/3} ~,
\label{eq_cascade}
\end{equation}
in which the dimensionless coefficient $A$ is a universal constant close to unity \citep{Kundu2012} and ignored hereafter.  
The value of $d$ ranges from the largest scale of 
the flow, which we denote $L$, down to the shortest eddy scale, which we denote $\ell$ and 
which is called the {\it Kolmogorov microscale}.
Applying \eqref{eq_cascade} at both extreme scales, we obtain:
\begin{itemize}
\item At the largest scale $L$ of the flow, the corresponding velocity scale is $V$.  Thus, 
by virtue of \eqref{eq_cascade}, we have $V \simeq (\epsilon L)^{1/3}$, from which we can 
extract the energy dissipation:
\begin{equation}
\epsilon \simeq \frac{V^3}{L} ~.
\label{eq_epsilon}
\end{equation}

\item At the shortest scale $\ell$, with associated minimum turbulent velocity 
$u_{min} \simeq (\epsilon \ell)^{1/3}$, molecular viscosity takes over because the Reynolds 
number at that scale becomes on the order of unity: $Re = \frac{\rho u_{min} \ell}{\mu} 
   = \frac{\rho (\epsilon \ell)^{1/3} \ell}{\mu}
   \simeq 1$.  This implies
\begin{equation}
\frac{\ell}{L} \simeq Re^{-3/4} ~.
\label{eq_ell_5}
\end{equation}   

\end{itemize}

For a high-Reynolds-number flow, \eqref{eq_ell_5} certifies that indeed $\ell \ll L$.

%----------------------------------------------------------------------------------------------------
%----------------------------------------------------------------------------------------------------
%----------------------------------------------------------------------------------------------------
%----------------------------------------------------------------------------------------------------

\section{Alternative Form of the Turbulent Stress Tensor}
\label{sec_alternate_tau_turb}
The stress \eqref{eq_tau_5L} 
\be
\tau_{ij}^{turb}{\ss(\x)} =   \rho \tfrac{(U\tau)^\alfa}{\tau}  (\alpha + 3) \Gamma(\alfa) \bar{C}_\alfa    \underset{\!\!\!\! |\x'-\x|  \geq \ell}{\iiint}    (x_i'-x_i) (x_j'-x_j)  \,      \frac{ (\x'-\x)  \cdot (\ub{\ss(\x')} - \ub{\ss(\x)})}{|\x'-\x|^{\alfa+5} }   \,d\x'  
\tag{\ref{eq_tau_5L}}
\ee
can be simplified by integrating by parts: 
$ \int a \, db = ab - \int b \, da $.  
First, write the dot product out into three terms 
$(x_k'-x_k) (\bu_k{\ss(\x')} - \bu_k{\ss(\x)})$, with implied summation over $k=1,2,3$.  Now, for each of the three terms, perform partial integration in $dx'_k$ 
\begin{align*}
a  &=  (x_i'-x_i) (x_j'-x_j) (\bu_k{\ss(\x')} - \bu_k{\ss(\x)})  \\
da &= \Big\{  \delta_{ik}(x_j'-x_j) (\bar{u}_k{\ss(\x')} - \bar{u}_k{\ss(\x)})
 			 + \delta_{jk}(x_i'-x_i) (\bar{u}_k{\ss(\x')} - \bar{u}_k{\ss(\x)})  \\
&\quad\quad  + (x_i'-x_i) (x_j'-x_j) \tfrac{\p \bu_k{\ss(\x')}}{\p x_k'} \Big\} dx_k'  \\
db &= \frac{ (x_k'-x_k) }{|\x'-\x|^{\alfa+5}}  \,dx_k' \\
b &=  \frac{  -1 }{(\alfa+3) |\x'-\x|^{\alfa+3}} 
\end{align*}
with no summation implied in each partial integration (fixed $k$).  
Since $ab\big|_{-\infty}^\infty = 0$ for each of these partial integrations (so long as $|\ub{\ss(\x')}| / |\x'|^{\alfa+1} \ra 0 $ as $|\x'| \ra \infty$), the other terms can be recombined (by now implying summation over $k$):
\begin{equation}
\begin{split}
\tau_{ij}^{turb}{\ss(\x)} 
= \rho & \tfrac{(U\tau)^\alfa}{\tau}  \Gamma(\alfa) \bar{C}_\alfa      \underset{\!\!\!\! |\x'-\x|  \geq \ell}{\iiint}    
   \left\{  \frac{ 1}{|\x'-\x|^{\alfa+3}}  \right\} 
  \\
  & \cdot
\Big\{   \delta_{ik}(x_j'-x_j) (\bar{u}_k{\ss(\x')} - \bar{u}_k{\ss(\x)}) 
 	   + \delta_{jk}(x_i'-x_i) (\bar{u}_k{\ss(\x')} - \bar{u}_k{\ss(\x)})   \\
  &
	 \quad\quad\quad\quad\quad\quad\quad\quad\quad\quad\quad\quad\quad\quad   + (x_i'-x_i) (x_j'-x_j) \tfrac{\p \bu_k{\ss(\x')}}{\p x_k'} \Big\} d\x'  ~.
\label{eq_tau_6L} 
\end{split}
\end{equation}
Assuming incompressible flow, $\tfrac{\p \bu_k}{\p x_k'} = 0$,  we are left with 
\be
\boxed{
\tau_{ij}^{turb}{\ss(\x)} =  \rho \tfrac{(U\tau)^\alfa}{\tau}   \Gamma(\alfa) \bar{C}_\alfa   \underset{\!\!\!\! |\x'-\x|  \geq \ell}{\iiint}    
\frac{  (x_j'-x_j) (\bar{u}_i{\ss(\x')} - \bar{u}_i{\ss(\x)}) + (x_i'-x_i) (\bar{u}_j{\ss(\x')} - \bar{u}_j{\ss(\x)}) }{|\x'-\x|^{\alpha + 3}}   d\x'
} ~.
\label{eq_tau_7L} 
\ee
It is easy to check that the friction force \eqref{eq_forcing_fractional_Laplacian} can alternatively be derived by taking the gradient of \eqref{eq_tau_7L}.

In the special case of the Cauchy distribution ($\alfa=1$, $U= \gamma\, u_*$), the stress \eqref{eq_tau_7L} becomes 
\be
\tau_{ij}^{turb}{\ss(\x)} =  \frac{\rho  \gamma\, u_*}{\pi^2}  \underset{\!\!\!\! |\x'-\x|  \geq \ell}{\iiint}    
\frac{  (x_j'-x_j) (\bar{u}_i{\ss(\x')} - \bar{u}_i{\ss(\x)}) + (x_i'-x_i) (\bar{u}_j{\ss(\x')} - \bar{u}_j{\ss(\x)}) }{|\x'-\x|^4}   d\x'  ~.
\label{eq_tau_7L_cauchy} 
\ee

%----------------------------------------------------------------------------------------------------
%----------------------------------------------------------------------------------------------------
%----------------------------------------------------------------------------------------------------
%----------------------------------------------------------------------------------------------------
\section{Reduction of Order of the Fractional Laplacian}
\label{sec_order_reduction}

The {\it fractional Laplacian} of scalar function $\bu{\ss(\x)}$ of vector $\x$ in $\mathbb{R}^n$ space is defined as \citep{Kwasnicki2017}:
\be
{\cal L}_n u{\ss(\x)} \equiv L_{\alfa,n}      
\dashint\!\!\dashint\!\!\dashint_{\!-\infty}^\infty 
\quad    \frac{  \bar{u}{\ss(\x')} - \bar{u}{\ss(\x)}  }{|\x'-\x|^{\alfa+n}} 
\,d\x' ~,
\quad \quad  \quad
L_{\alfa,n} \equiv \frac{2^\alfa \, \Gamma\(\frac{\alfa + n}{2}\)}{\pi^\frac{n}{2} \big|\Gamma\(-\frac{\alfa}{2}\)\big|} ~.
\label{eq_fractional_Laplacian}
\ee
Note that the coefficient in the fractional Laplacian, $L_{\alfa,n}$ is identical to the constant describing the tail of the L\'evy $\alfa$-stable distribution, $C_{\alfa,n}$.

Consider the case of $\x \in \mathbb{R}^n$ where scalar function $\bu{\ss(\x)}$ is not a function of $x_n$.  In this case, we expect that we can simply apply the definition of the fractional Laplacian  \eqref{eq_fractional_Laplacian} on the $\mathbb{R}^{n-1}$ subspace spanned by the components of $\x$ that $\bu{\ss(\x)}$ actually depends on.  Equivalently, we expect that integration over $dx_n$ in  ${\cal L}_n \bu{\ss(\x)}$ should leave remaining integrals identical to the formula for ${\cal L}_{n-1} \bu{\ss([x_1,\ldots,x_{n-1}])}$.  Using this thought experiment, we can derive the relation between constants $L_{\alfa,n}$ and $L_{\alfa,n-1}$.  Whence, we expect
\be
\begin{split}
\dashint_{\!-\infty}^\infty 
& \,\, \frac{ L_{\alfa,n} \,  dx_n'  }{[ (x_1-x_1')^2  + (x_2-x_2')^2 + \dots + (x_n-x_n')^2]^{\frac{\alfa+n}{2}}}   
\\
&\quad\quad\quad\quad\quad\quad\quad\quad\quad\quad =
 \frac{  L_{\alfa,n-1} }{[  (x_1-x_1')^2 + \dots + (x_{n-1}-x_{n-1}')^2]^{\frac{\alfa+n-1}{2}}} ~.
\end{split}
\label{eq_Ln_eq_Lnmo}
\ee
Upon setting $a^2 = (x_1-x_1')^2 + \dots + (x_{n-1}-x_{n-1}')^2$, $b^2 = (x_n-x_n')^2$, $b = a\tan\theta$, we have $a^2 + b^2 = a^2 (1+\tan^2\theta) = a^2 / \cos^2\theta$ and $db = a \, d\theta / \cos^2\theta$, such that the left hand side of \eqref{eq_Ln_eq_Lnmo} evaluates to 
\begin{align}
\dashint_{\!-\infty}^\infty  \,\, \frac{ L_{\alfa,n} \,  db  }{[ a^2 + b^2]^{\frac{\alfa+n}{2}}}  
&=
\dashint_{\!-\pi/2}^{\pi/2}  \,\, \frac{ L_{\alfa,n} \,  a \, d\theta / \cos^2\theta  }{[ a^2 / \cos^2\theta ]^{\frac{\alfa+n}{2}}}  
=
\frac{ L_{\alfa,n} }{[ a^2 ]^{\frac{\alfa+n-1}{2}}}  \sqrt{\pi}\frac{\Gamma\(\frac{\alfa+n-1}{2}\)}{\Gamma\(\frac{\alfa+n}{2}\)}  ~.
\end{align}
The right hand side of \eqref{eq_Ln_eq_Lnmo} is $\frac{  L_{\alfa,n-1} }{[a^2]^{\frac{\alfa+n-1}{2}}}$.  For these to be equal, we require
\be
 L_{\alfa,n-1} =  L_{\alfa,n}   \sqrt{\pi}\frac{\Gamma\(\frac{\alfa+n-1}{2}\)}{\Gamma\(\frac{\alfa+n}{2}\)} ~.
\ee
This requirement is consistent with the definition of $ L_{\alfa,n} $ in  \eqref{eq_fractional_Laplacian}.  So to recapitulate, if  scalar function $\bu{\ss(\x)}$ is not a function of $x_n$, that variable can be integrated out of the fractional Laplacian by simply applying the definition \eqref{eq_fractional_Laplacian} in the $\mathbb{R}^{n-1}$ subspace spanned by the remaining $n-1$ variables.  Moreover, if $\bu{\ss(\x)}$ only depends on $k$ elements of $\x$, then the fractional Laplacian can be applied only on that $\mathbb{R}^{n-k}$ subspace.  This is useful when we consider shear flows such as the Couette problem, since the velocity field has only $z$ dependency, $\bu = \bu{\ss(z)}$.

%===============================================================================
%     References:
%===============================================================================
\bibliographystyle{apa}
\bibliography{Turbulence.bib}

\end{document}